\documentclass[preprint,showpacs,preprintnumbers,amsmath,amssymb]{revtex4}

\usepackage{graphicx}% Include figure files
\usepackage{dcolumn}% Align table columns on decimal point
\usepackage{bm}% bold math

\begin{document}

%\preprint{APS/123-QED}

\title{Electronic transport in DNA functionalized graphene sensors}
\author{P. Gurung}
 %\altaffiliation[Also at ]{Physics Department, Delhi University.}%Lines break automatically or can be forced with \\
\author{N. Deo}%
\email{ndeo@physics.du.ac.in}
\affiliation{%
Department of Physics and Astrophysics, University of Delhi,
Delhi 110007, India.
}%


\begin{abstract}
A theoretical understanding of the experimental electronic transport phenomena
in gas sensors based on DNA functionalized graphene is presented by quantitatively
investigating the time-dependent electronic transport in these devices using the
nonequilibrium Green's function (NEGF) formalism and tight-binding approximation.
%A simple example to motivate the time dependence introduced in the hopping integral
%and on-site energy is given.
The time-dependent zeroth and first order contributions to the current are calculated
with derivations of the equation of motion and Dyson equation. The zeroth order
contribution is identified as the time-dependent Landauer formula in terms of the slow
time variable and the first order contribution is found to be small in this experiment.
The current is explicitly calculated by deriving a formula for the transmission function
and considering a form for the hopping integral which includes the effect of chemical
vapors on the charge distribution of the carbon atoms and the nearest-neighbor carbon-carbon
distance $\rm a_{cc}$. Theoretical results are found in agreement with the experimental
results. A shift in the Fermi level ($\rm \varepsilon_f$) is calculated, which is a result
of shift in the Dirac point due to adsorption of vapors on the DNA functionalized graphene.
%Using a set of values of $\rm \Delta a_{cc}$ and $\rm \varepsilon_f$ for a specific vapor and DNA
%sequence theoretical values of the current response can be predicted for different sequences.
The work suggests that using the same values of change in $\rm a_{cc}$ due to the four DNA
bases for a specific target vapor, the theoretical values of the current response can be predicted
for different DNA sequences leading to the application of the graphene sensors as a DNA analyser.
\end{abstract}


\maketitle

\section{Introduction}
Graphene is a single atomic layer of graphite that is truly a two-dimensional
structure. Graphene was first isolated in 2004 by micromechanical cleavage of
graphite \cite{1}. The exceptional and fascinating mechanical, structural and
electronic properties of graphene make it a promising building block for
next-generation electronics. The most remarkable property of graphene originates
from its unusual electronic structure. Graphene has a unique electronic band
structure where the valence and conduction bands meet at the Dirac points,
giving rise to conical sections of the energy spectrum. Hence, charge carriers
in graphene possess a linear dispersion relation similar to that described by
the massless Dirac equation and thus behave in a unique manner. They move as if they
have zero effective mass and a velocity which is about 300 times slower than
the speed of light \cite{2,3,4,5,6,7,8,9,10}. Graphene reveals ambipolar electric
field-effect where the charge carriers can be electrons or holes with their density
proportional to gate voltage \cite{1,2,10}. By changing
the gate voltage one can modify the charge carrier density in graphene and therefore
the Fermi level can be tuned. In pristine graphene, the Fermi level lies at the
Dirac point and generates zero density of states, that is, there are no states to
occupy and hence there are  no charge carriers which contribute to the electronic
transport, resulting in a large resistivity. The Dirac point separates the region
of conduction by electrons from the region where the transport is governed by holes.
When the Fermi level shifts away from the Dirac point (into hole or electron conduction
regime) on application of the gate voltage, the resistance decreases with increasing
gate voltage. As a result, the carrier concentration increases which contributes to
the transport. Graphene possesses extremely high carrier mobility, a measure of how
fast charges travel \cite{1,2,9,10,11,12,13,14}, leading to ballistic transport
\cite{1,2,12}, and also exhibits quantum hall effect \cite{3,10,15,16,17,18,19}.

Graphene has a large surface area \cite{20} with every carbon atom on its surface in
direct contact with surrounding atmosphere, making its electronic and mechanical
properties extremely sensitive to small changes in the environment. Hence, enabling
the fabrication of miniaturized sensors capable of detecting a number of molecular
species \cite{21,22,23,24,25,26,27,28,29} with high sensitivity, excellent selectivity,
and fast response and recovery time. Substantial progress in carbon nanotube (CNT)
based chemical sensors has already been achieved as addressed in the literatures
\cite{30,31,32,33,34,35,36}. But a major difficulty in such sensors is the separation
of semiconducting carbon nanotubes from metallic ones, while the more consistent
electronic structure of graphene generally makes it more suitable as compared to
CNTs for use in such devices. The sensing properties of graphene are found to be
promising for gases such as $\rm NH_3$,$\rm CO$,$\rm H_2O$, $\rm O_2$, $\rm NO_2$
and 2,4-dinitrotoluene (DNT) \cite{21,22,23,24,25,26,27,28,29}. The gas sensing
characteristics of graphene have been attributed to its two-dimensional structure
and the extraordinary mobility of charge carriers \cite{21,24}. Since graphene is
all surface, no bulk, thus there is a huge scope for studying the surface dependent
gas sensing phenomenon on its surface. The large surface area gives very high
sensitivity and the high electron mobility gives ultra fast response times. Graphene
is superior because of its high electrical conductivity (even if there are few
carriers) as well as low noise properties \cite{2,3,11,21} which make the local
change in its electrical properties, due to adsorption of gas molecules on its
surface acting as donors or acceptors, detectable.

Graphene has shown its potential for detection of various molecular species with
high sensitivity in ambient conditions. However, its performance is limited to
those molecular species that get adsorbed on the surface causing a detectable
sensor response. But a weak or no observable response of intrinsic graphene
(without resist residue on the surface) is found for water vapor, TMA, octanoic
acid \cite{23} and for DMMP and PA \cite{37}. To overcome this limitation of
intrinsic graphene, functionalization is a promising approach to tune its electronic
properties. The effect of functionalization of graphene with metal nanoparticles
and bio-molecules on their sensing properties in terms of enhanced reactivity and
sensitivity towards molecular species has already been reported \cite{37,38,39,40,41,42}.

Despite tremendous experimental progress in gas sensing applications of graphene
the underlying mechanism of gas adsorption and detection is not very clear and
needs an in-depth understanding. The path breaking experimental results on
functionalized graphene have raised challenging issues for theory and modeling,
especially, on how to describe the dynamics in the presence of dopants for electronic
transport measurements. To address this issue a quantitative and detailed understanding
of the electronic transport in such nanohybrids is needed.

This paper reports on the theoretical understanding of the experiment where the
electronic transport properties (sensing response) of devices based on graphene
monolayer FETs decorated with single stranded DNA (ssDNA) strands were measured
in the presence of chemical vapors \cite{37} for which the clean device does not
show any response. The electronic transport through the system is modeled using
the tight-binding Hamiltonian and the nonequilibrium Green's function (NEGF)
formalism. In the experiment reported in Ref. [37], the ssDNA oligomers were
applied to the surface of the graphene by putting a small drop of dilute solution
of ssDNA in deionized water (700 $\rm \mu g/mL$) onto the chip surface with the
graphene device \cite{37}. The device was then immediately placed into a chamber
held at 100\% relative humidity by a steam bath to prevent the solution from
evaporating off the surface. The sample was left in the humid chamber for 45 minutes,
after which the ssDNA solution was blown off with nitrogen gas. This process enabled
the self-assembly of a nanoscale ssDNA layer on the graphene surface. Two ssDNA
sequences, sequence 1: $5^\prime$ GAG TCT GTG GAG GAG GTA GTC $3^\prime$ and sequence
2: $5^\prime$ CTT CTG TCT TGA TGT TTG TCA AAC $3^\prime$ were selected to functionalize
the graphene device.

The vapor response measurements were recorded at 1 mV applied bias voltage and zero
gate voltage with varying concentrations. Initially,
the carrier gas ($\rm N_2$) was passed through the chamber at a rate of 1 slm and
then chemical vapor dimethylmethylphosphonate (DMMP) was substituted for a small
percentage of the $\rm N_2$ flow with the total flow rate held constant \cite{37}.
Figure 2(a) of Ref. [37] shows the responses to DMMP vapor of devices based on
pristine graphene, graphene functionalized with ssDNA sequence 1, and ssDNA sequence
2 at different concentrations. The sensor responses are presented as changes in the
device current normalized to the current measured when pure $\rm N_2$ flowed. The
current response of clean graphene to DMMP vapor is very weak and barely detectable
above the noise level, at all concentrations, although a response $\Delta I/I_0\sim
1\%$ was recorded at the highest concentration. Functionalization of graphene with
ssDNA enhanced the responses on the scale of 5\% to 50\%. The sensor responses were
reproducible with nearly perfect recovery to baseline upon purging \cite{37}. Because
of the negatively charged backbone of ssDNA, its role is to make more hydrophilic
environment around the chemically inert and hydrophobic conduction channel of graphene
which facilitates the adsorption of more analytes on the device \cite{37}. As a result,
the sensor response increases compared to the response of clean graphene. The measured
sensor responses of functionalized graphene device for DMMP with sequence 1 and 2 are
given in Table 1 of Ref. [37].

\section{Tight-binding model and nonequilibrium Green's function}
\label{2}
To study the electronic transport in the DNA-decorated graphene-FET, a simplified
theoretical model of the complex Gas-DNA-graphene-FET system is presented. The model
consists of a graphene monolayer, acting as the channel, connected to two metallic
contact electrodes source (S) and drain (D) with a single strand of ssDNA sequence
1 applied on its surface as shown in Fig. 1. The model represents a graphene sheet
with carbon atoms labeled as $\rm {A_1, B_1, A_2, B_2,\cdots X}_{i}$, where X can be
either carbon atom A or B and the index $\rm i=1, 2, \cdots, \rm M/2\rm ~or~ (M+1)/2$
denotes the number of the carbon atoms, and M is the total number of carbon atoms
involved in the transport. Since graphene is decorated with two ssDNA sequences,
sequence 1 with 21 bases and sequence 2 with 24 bases. Therefore the value of M can
be odd or even depending on the type of the sequence. For ssDNA sequence 1, M is odd
and is equal to 23, while M is even and equal to 26 for sequence 2 including the first
$\rm A_1$ and the last $\rm X_{M/2}(X_{(M+1)/2})$ carbon atoms which are connected to
S and D electrodes. In the present model, the decoration of graphene is shown for ssDNA
sequence 1 with bases guanine (G), adenine (A), thymine (T) and cytosine (C), in the
order of arrangement GAG TCT GTG GAG GAG GTA GTC, and denoted by green ovals interacting
with different carbon atoms and the pink and purple circles represent the DMMP vapor
and $\rm N_2$ gas, Fig. 1. %The arrow indicates the path of transmission in Fig. 7.6.

In a graphene sheet, the carbon atoms are held together by $\rm sp^2$ hybridized
covalent bonds, while the electronic transport takes place by hopping along $\pi$
orbitals which can participate in some kind of bonding with adsorbates \cite{43}.
Here, it is assumed that the bonding is a kind of van der Waals (vdW) bonding
($\pi-\pi$ stacking) between the DNA bases and the carbon atoms.

%The working of the model is based on the changes in its
The electrical properties of the graphene device changes as a result of adsorption
of the DNA bases and vapor molecules on the graphene surface. The experimental sensor
response is found to be sequence sensitive. Therefore, initially, when the graphene
is decorated with ssDNA sequence 1, the base guanine (G) interacts with the carbon
atom $\rm B_1$ and the adenine (A) base interacts with $\rm A_2$ and the rest of the
bases keep on interacting with other carbon atoms. Upon DMMP exposure, the vapor
molecule interacts with G base at an instant of time $t_1$ through vdW forces,
Fig. 1(a). As a result, the electron hops from the $\pi$ orbital of the carbon atom
$\rm A_1$ to the neighboring carbon atom $\rm B_1$ and the tunneling of the electron
from $\rm A_1$ to $\rm B_1$ is an elastic process with the corresponding integral
called the hopping integral $\gamma_{11}(t_1)$. Then, the electron hops from
$\rm B_1$ to $\rm A_2$ with the hopping integral $\gamma_{12}(t_1)$, and from
$\rm A_2$ to $\rm B_2$ with $\gamma_{22}(t_1)=\gamma_0$, where $\gamma_0$ is
the hopping integral of the ssDNA coated graphene when exposed to $\rm N_2$ (without
the vapor). Hence, the hopping integrals between the rest of the carbon atoms
$\gamma_{\rm ij}(t_1)$ are also $\gamma_0$ as shown in Fig. 1(a). The indices
$\rm i(j)$=1,2,$\cdots {\rm p(q)}$, where $\rm p=q= M/2$ for even M and p= {(M-1)}/2
and q={(M+1)}/2 for odd M. In a similar manner, at time $t_{\rm (M+1)/2}$ (for
odd M) the time at which the vapor molecule interacts with the last base cytosine (C)
of the DNA sequence 1 attached to the carbon atom, the hopping integrals are $\gamma_{11}
(t_{(\rm M+1)/2}),\gamma_{12}(t_{(\rm M+1)/2}), \cdots \gamma_{\rm pq}(t_{
(\rm M+1)/2})$, Fig. 1(b). Hence, as a result of time-dependent interaction of vapor
molecules with the DNA-decorated graphene, the hopping integrals as well as the on-site
energy change with the time $t$ at which the vapor molecules trigger the different
bases of the DNA sequence attached to the graphene. This time-dependent hopping integral
and on-site energy give rise to time-dependent Hamiltonian and Green's function for the
functionalized graphene. Thus, to study the time-dependent electronic transport through
the model system the time-dependent NEGF formalism is well suited.

\begin{figure}[h!]
\begin{center}
\includegraphics[width=7cm]{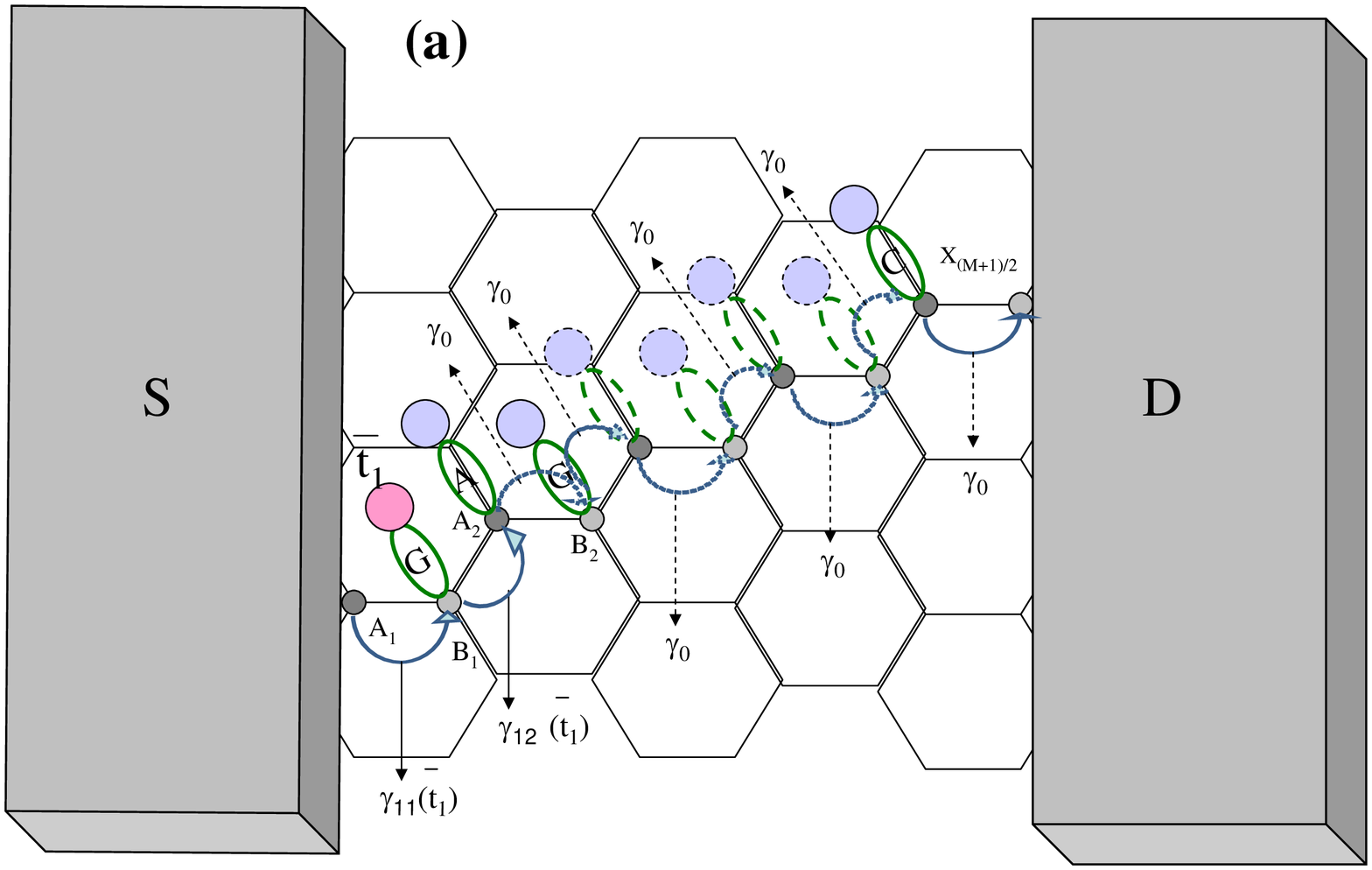}
\includegraphics[width=7cm]{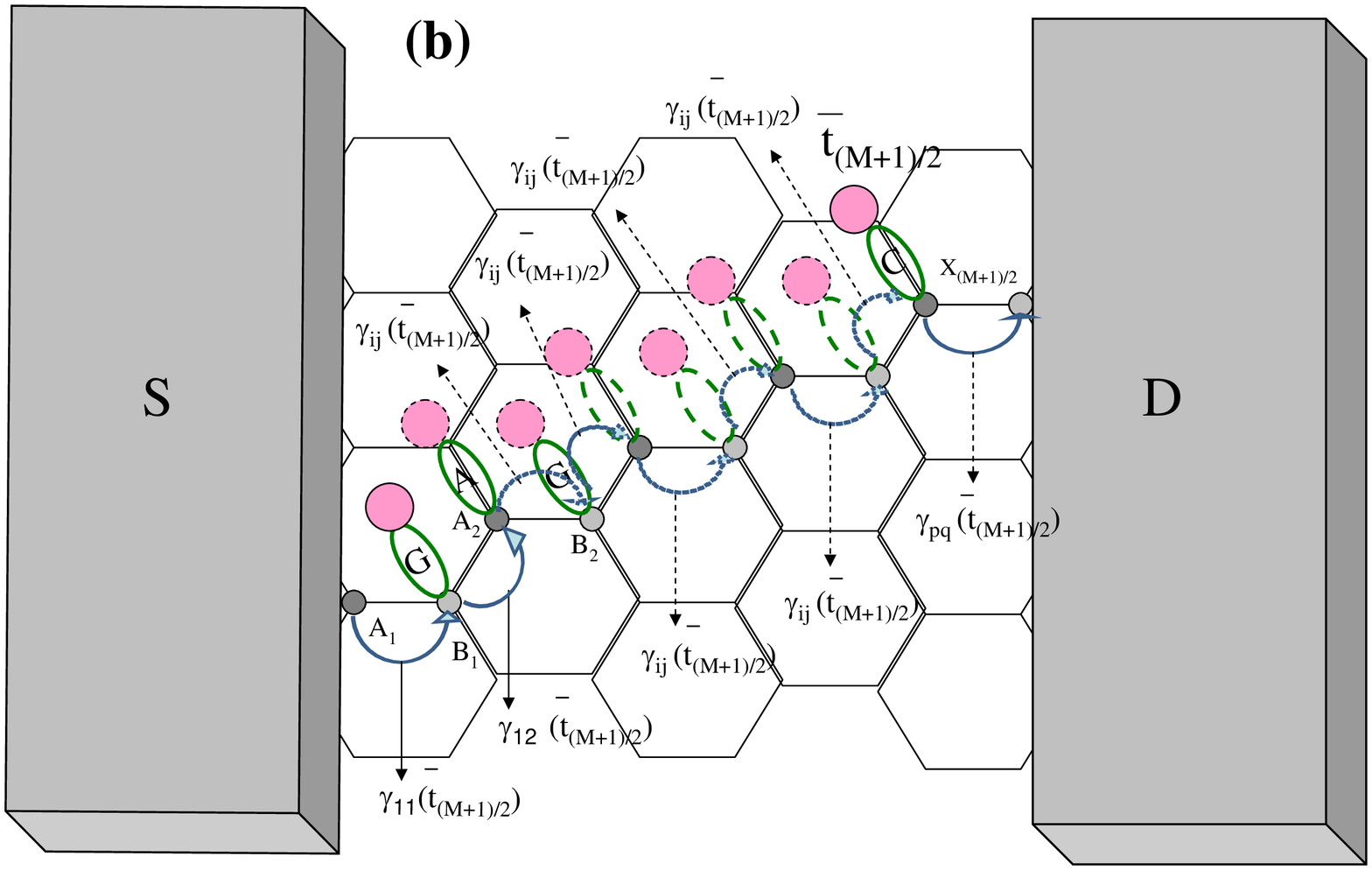}
\vspace{1cm} \caption{Adsorption of DMMP vapor on graphene decorated with ssDNA sequence
1 which is connected to $\rm S$ and $\rm D$ electrodes. The DNA bases are represented by
green ovals, DMMP and $\rm N_2$ molecules are indicated by pink and purple spheres. The
arrow indicates the path of transmission. ${\gamma_{\rm ij}(t)}$ are the hopping
integrals between different carbon atoms $\rm A_i$ and $\rm B_i$ at time $t$. Here,
$t=\bar{t}_1, \cdots, \bar{t}_{(\rm M+1)/2}$ after applying the adiabatic approximation (subsection
IIB) and $\gamma_0$ represents the hopping integral in the absence of DMMP.}
\label{fig.7.3}
\end{center}
\end{figure}

In order to understand the sensor response of the graphene device a simple model and
its working is presented because the actual experimental picture of the system is
quiet complicated. Graphene device is functionalized with two ssDNA sequences 1 and
2. Both sequences consist of four types of nucleobases G, C, A and T, respectively,
but with different order of arrangement of these four bases. In the experiment \cite{37},
there are many strands of the same ssDNA sequence either 1 or 2 applied on the surface
of the graphene sheet. Hence one expects an average sequence-independent effect, but
this is not observed in the experiment. Experimental responses of devices to DMMP vapor
are found to be sequence-dependent, the response of devices for sequence 1 is different
than that for sequence 2. Thus in order to show the sequence-dependent effect as observed
in the experiment, the model simply assumes the effect to be due to different charge
transfer by different bases of a single strand of ssDNA sequence of type either 1 or 2.

The Hamiltonian corresponding to the coupled Gas-DNA-graphene-FET system is given as

\begin{equation}
H_{\rm total}=H_{\rm graphene}+H_{\rm contact}+H_{\rm tunneling},
\end{equation}

where the Hamiltonian of graphene is expressed as
\begin{equation}
H_{\rm graphene}=\sum_m \varepsilon_m(t)d_m^\dagger d_m+\sum_{<mn>}\gamma_{mn}(t)d_m^\dagger d_n,
\end{equation}

where $d_m^\dagger$ and $d_m$ are creation and annihilation operators of an electron
in graphene, $\varepsilon_m (t)$ is the on-site energy of the carbon atom and $\gamma_{mn}
(t)$ is the nearest-neighbor hopping integral between the carbon atoms as a function
of time.

The contact Hamiltonian is written as
\begin{equation}
H_{\rm contact}=\sum_{k\alpha\epsilon {\rm S, D}}
\varepsilon_{k\alpha}c^\dagger_{k\alpha} c_{k\alpha},
\end{equation}
where $c^\dagger_{k\alpha}$ and $c_{k\alpha}$ are creation and annihilation operators
for  electrons with momentum $k$ in the contact, either S or D. In the experiment,
there is no external time-dependent bias applied between the S and D contacts therefore
the contact Hamiltonian is time-independent and electrons in the contacts are
non-interacting. The tunneling Hamiltonian which describes the coupling between the
contact electrodes and graphene in the absence of a time-dependent gate voltage is
expressed as
\begin{equation}
H_{\rm tunneling}=H_T= \sum_{n;k\alpha\epsilon {\rm S,D}}
(V_{k\alpha, n} c^\dagger_{k\alpha} d_n + H.c.).
\end{equation}

In the matrix form, the total Hamiltonian is expressed as
\begin{equation}
H(t) = \left(\begin{array}{ccc}
H_{\rm S} & V_{\rm {Sgrp}} & 0  \\
V^\dagger_{\rm {Sgrp}} & H_{\rm grp}(t) & V_{\rm {grpD}} \\
0 & V^\dagger_{\rm {grpD}} & H_{\rm D} \\
\end{array} \right),
\end{equation}
where $H_{\{\rm S,\rm D\}}$ are the Hamiltonians of the S and D contact electrodes
(Eq. (3)), and $H_{\rm grp}$ is the time-dependent tight-binding Hamiltonian for
graphene (Eq. (2)) with matrix elements defined as $H_{\rm A_iA_i}$=${\varepsilon_{
\rm Ai}(t)}$, $H_{\rm B_iB_i}$=${\varepsilon_{\rm Bi}(t)}$, the on-site
energy and $H_{\rm A_iB_i}$=$H_{\rm B_iA_i}$=${\gamma_{\rm ii}(t)}$ \cite{44},
$H_{\rm B_iA_{i+1}}$=$H_{\rm A_{i+1}B_i}$=${\gamma_{\rm i,i+1}(t)}$, the
hopping integrals and the rest of the off diagonal elements are zero. $V_{\rm {Sgrp}}$
and $V_{\rm{grpD}}$ are the coupling (tunneling) matrices between the two electrodes
and graphene (Eq. (4)). Hence, the time-dependent tight-binding Hamiltonian for
graphene is expressed as

$H_{\rm grp}(t)=$
\begin{eqnarray}
%G^{(0) r}_{\rm C}(\varepsilon,\bar{t})&=&\nonumber\\
%H_{\rm grp}(t)=
\left(
\begin{array}{ccccccccc}
\varepsilon_{\rm A_{1}}(t) & {\gamma_{11}(t)} &  &  &&\\
{\gamma_{11}(t)} & {\varepsilon_{\rm B_{1}}(t)} & {\gamma_{12}(t)} & &&\\
& {\gamma_{12}(t)} & {\varepsilon_{\rm A_{2}}(t)} & \ddots&& \\
&& \ddots & \ddots &\ddots &\\
&&&\ddots&\varepsilon_{\rm B_4}(t)&\gamma_{\rm pq}(t)\\
&&&&\gamma_{\rm pq}(t)&\varepsilon_{\rm X_i}(t)
\end{array} \right).
\end{eqnarray}

The effect of DMMP vapor and DNA is included in the graphene Hamiltonian in the form
of time-dependent on-site energy and hopping integrals because as the time changes,
the number of vapor molecules and hence their interaction with different DNA bases
changes, which results in a change in the on-site energy and hopping integral leading
to a change in sensor response. The reason for this tight-binding Hamiltonian can be
understood with an example of a single adsorbate interacting with graphene as described
in Ref. [43] where chemically adsorbed molecules are incorporated into the tight-binding
model of graphene as laterally attached additional sites, where the on-site and coupling
energies are extracted from the band structure of a graphene sheet with regularly placed
adsorbates. But in the present model system, the adsorbed molecules are included in the
tight-binding Hamiltonian not as attached additional sites. Here, the effect of the
molecules are incorporated into the change in the on-site energy and hopping between
the adjacent carbon sites as a function of interaction time of DMMP molecules with
the DNA coated graphene. Thus the effect of interaction of DMMP molecules on $\gamma_0$
and $\varepsilon_0$ of the pristine graphene is represented by taking a time-dependent
Hamiltonian $H(t)$ with $\gamma_0$ $\rightarrow $$\gamma(t)$ and $\varepsilon_0\rightarrow
\varepsilon(t)$. This indicates that as the time changes, the number of adsorbates and
hence the interaction changes, which results in a change in $\gamma_0$, $\varepsilon_0$
and the sensor response.

\subsection{Expression for the current}
\label{3}
%To investigate the electronic transport through the model system a general expression
%for the time-dependent current is derived using the standard technique of the NEGF
%formalism \cite{45,46,47,48,49,50,51,52,53} which is modified to fit the present model
%system, Fig. 1.
In the present model system, the time-dependence arises because of the interaction of
DMMP molecules with the DNA-graphene system for a given vapor exposure time not because
of externally applied time-dependent bias between the two electrodes, and a time-dependent
gate voltage.
%To investigate the electronic transport through the model system a general expression
%for the time-dependent current is derived using the standard technique of the NEGF
%formalism \cite{45,46,47,48,49,50,51,52,53} which is modified to fit the present model
%system, Fig. 1.
Using the standard technique of the NEGF formalism \cite{45,46,47,48,49,50,51,52,53} which
is modified to fit the present model system,
%and following the steps given in Ref. [], the equation of motion
%for the Green’s function is derived for the present model system. Hence,
the equation of motion for the Green’s function is derived and a general expression for
the time-dependent current flowing from the source or drain contact into the graphene device
is given as
\begin{equation}
I_{\rm S/D}(t)=\frac{-2e}{\hbar}\int^t_{-\infty} dt^\prime\int\frac{d\varepsilon}{2\pi}{\rm Im Tr} \bigg\{e^{-i\varepsilon_{k\alpha}(t^\prime-t)}{\Gamma}_{\rm S/D}(\varepsilon)\bigg[f_{\rm S/D}(\varepsilon){G}^r_{\rm grp}(t,t^\prime)+{G}^<_{\rm grp}(t,t^\prime)\bigg]\bigg\},
\end{equation}
%\end{widetext}

where $G_{\rm grp}$ denotes the Green's function of graphene in the presence of the
coupling with the contacts.

\subsection{Adiabatic approximation}
In the adiabatic approximation, the time scale over which the system parameters
change is large compared to the life time of an electron in the system (graphene)
\cite{55}. In the problem of electronic transport through the graphene-based sensor
the experimental time scale of the gas flow is much longer ($\sim \rm seconds$) than
the time scale of electron transport inside graphene ($\sim 10^{-15}~ \rm s$).
Therefore the adiabatic approximation is applied to study the electronic transport
in the gas sensor.

\subsubsection{Adiabatic expansion for the Green's functions}
The best way to separate slow and fast time scales is to re-parameterize the Green's
functions. In other words the time variables of the Green's functions are replaced
by a fast time difference $\delta t=t-t^{\prime}$ and a slow mean time
$\bar t=\frac{t+t^{\prime}}{2}$ as \cite {55}
\begin{equation}
G(t, t^\prime)=G\bigg(t-t^\prime, \frac{t+t^\prime}{2}\bigg).
\end{equation}

The adiabatic approximation is applied to lowest order by expanding the Green's
functions about the slow time variable up to linear order in the fast time variables
\cite{55}
\begin{eqnarray}
G(t-t^\prime,\bar t)&=&G(t-t^\prime, \bar t)|_{\bar t =t}+\Big(\frac{t^\prime-t}{2}\Big)\frac{\partial G}
{\partial \bar t}(t-t^\prime,\bar t)|_{\bar t=t}\nonumber\\
&=& G(t-t^\prime,t)+ \Bigg(\frac{t^\prime-t}{2}\Bigg)\frac{\partial G}{\partial \bar t}(t-t^\prime,\bar t)|_{\bar t=t},
%&=& G^{(0)}(t-t^\prime, \bar t)+G^{(1)}(t-t^\prime, \bar t),
\end{eqnarray}
which can be further written as
\begin{equation}
G(t-t^\prime,\bar t)= G^{(0)}(t-t^\prime, \bar t)+G^{(1)}(t-t^\prime, \bar t),
\end{equation}

where $G^{(0)}(t-t^\prime, \bar t)$ and $G^{(1)}(t-t^\prime, \bar t)$ are the zeroth
and first order Green's functions which lead to the zeroth and first order contributions
to the current.

Applying adiabatic expansion on $G(t,t^\prime)$ in Eq. (7) the current becomes

%\begin{widetext}
\begin{eqnarray}
I_{\rm S/D}(t-t^\prime,\bar t)&=&\frac{-2e}{\hbar}\int^t_{-\infty} dt^\prime\int\frac{d\varepsilon}{2\pi}{\rm Im Tr} \bigg\{e^{-i\varepsilon_{k\alpha}(t^\prime-t)}{\Gamma}_{\rm S/D}(\varepsilon)\bigg[f_{\rm S/D}(\varepsilon)G^r_{\rm grp}(t-t^\prime, \bar t)\nonumber\\
&&+G^<_{\rm grp}(t-t^\prime,\bar t)\bigg]\bigg\}\nonumber\\
&=&\frac{-2e}{\hbar}\int^t_{-\infty} dt^\prime\int\frac{d\varepsilon}{2\pi}{\rm Im Tr} \bigg\{e^{-i\varepsilon_{k\alpha}(t^\prime-t)}{\Gamma}_{\rm S/D}(\varepsilon)\bigg[f_{\rm S/D}(\varepsilon)\bigg({G}^{(0)r}_{\rm grp}(t-t^\prime, \bar t)\nonumber\\
&&+{G}^{(1)r}_{\rm grp}(t-t^\prime, \bar t)\bigg)+\bigg(G^{(0)<}_{\rm grp}(t-t^\prime,\bar t)+G^{(1)<}_{\rm grp}(t-t^\prime,\bar t)\bigg)\bigg]\bigg\}.
\end{eqnarray}
%\end{widetext}

Consider only the zeroth order contribution to the current

\begin{eqnarray}
I^{(0)}_{\rm S/D}(t-t^\prime,\bar t)&=&\frac{-2e}{\hbar}\int^t_{-\infty} dt^{\prime}\int\frac{d\varepsilon}{2\pi}{\rm Im Tr} \bigg\{e^{-i\varepsilon_{k\alpha}(t^\prime-t)}{\Gamma}_{\rm S/D}(\varepsilon)\bigg[G^{(0)<}_{\rm grp}
(t-t^\prime,\bar t)\nonumber\\
&&+ f_{\rm S/D}(\varepsilon){G}^{(0)r}_{\rm grp}(t-t^\prime, \bar t)\bigg]\bigg\}\nonumber\\
&=&I^{(0)}_1(t-t^\prime,\bar t)+I^{(0)}_2(t-t^\prime,\bar t).
\end{eqnarray}

%\begin{eqnarray}
%I^{(0)}_{\rm S/D}(t-t^\prime,\bar t)=&&\frac{-2e}{\hbar}\int^t_{-\infty} dt^{\prime}\int\frac{d\varepsilon}{2\pi}{\rm Im Tr} \bigg\{e^{-i\varepsilon_{k\alpha}(t^\prime-t)}\nonumber\\
%&&\times{\Gamma}_{\rm S/D}(\varepsilon)\bigg[G^{(0)<}_{\rm grp}
%(t-t^\prime,\bar t)+ f_{\rm S/D}(\varepsilon)\nonumber\\
%&&\times{G}^{(0)r}_{\rm grp}(t-t^\prime, \bar t)\bigg]\bigg\}\nonumber\\
%&&=I^{(0)}_1(t-t^\prime,\bar t)+I^{(0)}_2(t-t^\prime,\bar t).
%\end{eqnarray}

Solving for the first term %by taking the Fourier transform

\begin{eqnarray*}
I^{(0)}_1(t-t^\prime,\bar t)&=&\frac{-2e}{\hbar}\int^t_{-\infty} dt^{\prime}\int\frac{d\varepsilon}{2\pi}{\rm Im Tr} \bigg\{e^{-i\varepsilon_{k\alpha}(t^\prime-t)}{\Gamma}_{\rm S/D}(\varepsilon)G^{(0)<}_{\rm grp}(t-t^\prime,\bar t)\bigg\}.\nonumber\\
%&=&\frac{-2e}{\hbar}\int\frac{d\varepsilon}{2\pi}{\rm Im Tr} \bigg\{\int^t_{-\infty} dt^\prime e^{-i\varepsilon_{k\alpha}(t^{\prime}-t)}{\Gamma}_{\rm S/D}(\varepsilon)G^{(0)<}_{\rm grp}(t-t^{\prime},\bar t)\bigg\}\nonumber\\
\end{eqnarray*}

Rearranging the terms and changing variable $t^\prime$ to $(t-t^\prime)$ and the
limits we get

\begin{eqnarray*}
I^{(0)}_1(t-t^\prime,\bar t)&=& \frac{-2e}{\hbar}\int\frac{d\varepsilon}{2\pi}{\rm Im Tr} \bigg\{{\Gamma}_{\rm S/D}(\varepsilon)\int^\infty_{0} d(t-t^{\prime}) e^{i\varepsilon_{k\alpha}(t-t^{\prime})}G^{(0)<}_{\rm grp}(t-t^{\prime},\bar t)\bigg\},\nonumber\\
%\end{eqnarray*}
%\end{widetext}
%which is further written as
%\begin{widetext}
%\begin{equation*}
&=&\frac{-e}{\hbar}\int\frac{d\varepsilon}{2\pi}{\rm Im Tr} \bigg\{{\Gamma}_{\rm S/D}(\varepsilon)\int^\infty_{-\infty} d(t-t^{\prime}) e^{i\varepsilon_{k\alpha}(t-t^{\prime})}G^{(0)<}_{\rm grp}(t-t^{\prime},\bar t)\bigg\}.\nonumber\\
\end{eqnarray*}

%\begin{eqnarray*}
%I^{(0)}_1(t-t^\prime,\bar t)=&& \frac{-2e}{\hbar}\int\frac{d\varepsilon}{2\pi}{\rm Im Tr} \bigg\{{\Gamma}_{\rm S/D}(\varepsilon)\int^\infty_{0} d(t-t^{\prime})\nonumber\\
%&&\times e^{i\varepsilon_{k\alpha}(t-t^{\prime})}G^{(0)<}_{\rm grp}(t-t^{\prime},\bar t)\bigg\},\nonumber\\
%%\end{eqnarray*}
%%\end{widetext}
%%which is further written as
%%\begin{widetext}
%%\begin{equation*}
%&=&\frac{-e}{\hbar}\int\frac{d\varepsilon}{2\pi}{\rm Im Tr} \bigg\{{\Gamma}_{\rm S/D}(\varepsilon)\int^\infty_{-\infty} d(t-t^{\prime})\nonumber\\
%&&\times  e^{i\varepsilon_{k\alpha}(t-t^{\prime})}G^{(0)<}_{\rm grp}(t-t^{\prime},\bar t)\bigg\}.\nonumber\\
%\end{eqnarray*}

Taking the Fourier transform we obtain
%\begin{widetext}
%\begin{equation}
\[
I^{(0)}_1(\bar t)=\frac{-e}{\hbar}\int\frac{d\varepsilon}{2\pi}{\rm Im Tr} \bigg\{{\Gamma}_{\rm S/D}(\varepsilon)G^{(0)<}_{\rm grp}
(\varepsilon,\bar t)\bigg\}, %\nonumber\\
\]
%\end{equation}
%\end{widetext}
and using the definition of $G^<_{nm}(t,t^\prime)=\frac{i}{\hbar}<d^\dagger_m(t^\prime)d_n(t)>$
the first term is written as
%\begin{widetext}
\begin{equation}
I^{(0)}_1(\bar t)=\frac{ie}{\hbar}\int\frac{d\varepsilon}{2\pi}{\rm Tr}\bigg\{{\Gamma}_{\rm S/D}(\varepsilon){G}^{(0)<}_{\rm grp}
(\varepsilon,\bar t)\bigg\}. %~~~{\rm (using~ Eq. ~(22) ~for} ~G^<).
\end{equation}
%\end{widetext}

Now solving for the second term

\begin{eqnarray*}
I^{(0)}_2(t-t^\prime,\bar t)&=&\frac{-2e}{\hbar}\int^t_{-\infty} dt^{\prime}\int\frac{d\varepsilon}{2\pi}{\rm Im Tr} \bigg\{e^{-i\varepsilon_{k\alpha}(t^{\prime}-t)}{\Gamma}_{\rm S/D}(\varepsilon)f_{\rm S/D}(\varepsilon)
{G}^{(0)r}_{\rm grp}(t-t^{\prime}, \bar t)\bigg\}\nonumber\\
%&=&\frac{-2e}{\hbar}\int\frac{d\varepsilon}{2\pi}{\rm Im Tr} \bigg\{\int^t_{-\infty} dt^{\prime}e^{-i\varepsilon_{k\alpha}(t^{\prime}-t)}{\Gamma}_{\rm S/D}(\varepsilon)f_{\rm S/D}(\varepsilon){G}^{(0)r}_{\rm grp}(t^-t^{\prime}, \bar t)\bigg\}\nonumber\\
&=&\frac{-2e}{\hbar}\int\frac{d\varepsilon}{2\pi}{\rm Im Tr} \bigg\{\int^\infty_{0} d(t-t^{\prime})e^{i\varepsilon_{k\alpha}(t-t^{\prime})}{\Gamma}_{\rm S/D}(\varepsilon)f_{\rm S/D}(\varepsilon){G}^{(0)r}_{\rm grp}(t-t^{\prime}, \bar t)\bigg\}, \nonumber\\
\end{eqnarray*}

since ${G}^{(0)r}_{\rm grp}(t-t^{\prime}, \bar t)$=0 for $t-t^{\prime}<0$ from the definition
of $G^r_{nm}(t, t^\prime)=\frac{-i}{\hbar}\theta(t-t^\prime)<\{d_n(t),d^\dagger_m(t^\prime)\}>$,
therefore the second term becomes

\begin{eqnarray}
I^{(0)}_2(\bar t)&=&\frac{-2e}{\hbar}\int\frac{d\varepsilon}{2\pi}{\rm Im Tr}\bigg\{{\Gamma}_{\rm S/D}(\varepsilon)f_{\rm S/D}(\varepsilon)
{G}^{(0)r}_{\rm grp}(\varepsilon, \bar t)\bigg\}\nonumber\\
%&=&\frac{-2e}{\hbar}\int\frac{d\varepsilon}{2\pi}{\rm Tr}\bigg\{{\Gamma}_{\rm S/D}(\varepsilon)f_{\rm S/D}
%(\varepsilon)\bigg[\frac{{G}^{(0)r}_{\rm grp}(\varepsilon, \bar t)-{G}^{(0)a}_{\rm grp}(\varepsilon, \bar t)}{2i}\bigg]\bigg\}\nonumber\\
&=&\frac{ie}{\hbar}\int\frac{d\varepsilon}{2\pi}{\rm Tr}\bigg\{{\Gamma}_{\rm S/D}(\varepsilon)f_{\rm S/D}
(\varepsilon)\bigg[{G}^{(0)r}_{\rm grp}(\varepsilon, \bar t)-{G}^{(0)a}_{\rm grp}(\varepsilon, \bar t)\bigg]\bigg\}.
\end{eqnarray}

Putting Eqs. (13) and (14) in Eq. (12) we get
%\begin{widetext}
\begin{equation}
I^{(0)}_{\rm S/D}(\bar t)=\frac{ie}{\hbar}\int\frac{d\varepsilon}{2\pi}{\rm Tr}\bigg({\Gamma}_{\rm S/D}(\varepsilon)\bigg\{G^{(0)<}
_{\rm grp}(\varepsilon,\bar t)+f_{\rm S/D}(\varepsilon)\bigg[G^{(0)r}_{\rm grp}(\varepsilon, \bar t)-G^{(0)a}_{\rm grp}(\varepsilon, \bar t)\bigg]\bigg\}\bigg).
\end{equation}
%\end{widetext}

\subsection{Green's function for graphene}

The time-ordered Green's function of graphene $G_{nm}(t,t^\prime)$ satisfies the equation of motion
%\begin{eqnarray}
%%\lefteqn{
%\bigg[i\hbar\frac{\partial}{\partial t}-\varepsilon_n(t)\bigg]G_{nm}(t,t^\prime) &=&\delta (t-t^\prime)\delta_{nm}+\sum_{<n n^\prime>}\gamma_{n n^\prime}(t)G_{n^\prime m}(t,t^\prime)+\sum_{k^\prime\alpha^\prime}V^*_{k^\prime\alpha^\prime, n}G_{k^\prime\alpha^\prime,m}(t,t^\prime),
%\end{eqnarray}
%
%which can be further written as

\begin{equation}
%\lefteqn{
\bigg[i\hbar\frac{\partial}{\partial t}\delta_{nn^\prime}-\varepsilon_n(t)
\delta_{nn^\prime}-\sum_{<n n^\prime>}\gamma_{n n^\prime}(t)\bigg]G_{n^\prime m}(t,t^\prime)= \delta (t-t^\prime)\delta_{nm}+ \sum_{k^\prime\alpha^\prime}V^*_{k^\prime\alpha^\prime, n}G_{k^\prime\alpha^\prime,m}(t,t^\prime),
\end{equation}
%\end{widetext}

where $G_{k^\prime\alpha^\prime,m}(t,t^\prime)=g_{k^\prime\alpha^\prime}(t-t^\prime)\sum_{m^\prime}
V_{k^\prime\alpha^\prime,m^\prime}G_{m^\prime m}(t,t^\prime)$ is the contact time-ordered Green's function.

Equation (16) can be further written as
%Substituting $G_{k^\prime\alpha^\prime,m}(t,t^\prime)=g_{k^\prime\alpha^\prime}(t-t^\prime)
%\sum_m^\prime V_{k^\prime\alpha^\prime,m^\prime}\newline G_{m^\prime m}(t,t^\prime)$ from Eq. (16) in Eq.(13) one gets

\begin{eqnarray}
%\lefteqn{
\bigg[i\hbar\frac{\partial}{\partial t}\delta_{nn^\prime}-\varepsilon_n(t)\delta_{nn^\prime}-\sum_{<n n^\prime>}\gamma_{n n^\prime}(t)\bigg]G_{n^\prime m}(t,t^\prime)&=& \delta (t-t^\prime)\delta_{nm}+ \sum_{k^\prime\alpha^\prime}V^*_{k^\prime\alpha^\prime, n} g_{k^\prime\alpha^\prime}(t-t^\prime)\nonumber\\
&\times&\sum_{m^\prime} V_{k^\prime\alpha^\prime, m^\prime}G_{m^\prime m}(t,t^\prime)\nonumber\\
&=& \delta (t-t^\prime)\delta_{nm}+ \sum_{m^\prime}\Sigma_{nm^\prime}(t-t^\prime)G_{m^\prime m}(t,t^\prime),
\end{eqnarray}

where
\begin{equation}
\Sigma_{nm^\prime}(t-t^\prime)=\sum_{k^\prime\alpha^\prime\epsilon {\rm S,D}}V^*_{k^\prime\alpha^\prime,n}g_{k^\prime\alpha^\prime}(t-t^\prime)V_{k^\prime\alpha^\prime,m^\prime},
\end{equation}

is the time-independent self-energy \cite{51}.

After replacing $m^\prime$ with $n^\prime$ Eq. (17) can be further written as
%\begin{widetext}
\begin{equation}
%\lefteqn{
\bigg[i\hbar\frac{\partial}{\partial t}\delta_{nn^\prime}-\varepsilon_n(t)\delta_{nn^\prime}-\sum_{<n n^\prime>}\gamma_{n n^\prime}(t)\bigg]G_{n^\prime m}(t,t^\prime)=\delta (t-t^\prime)\delta_{nm}+ \Sigma_{nn^\prime}G_{n^\prime m}(t,t^\prime).
\end{equation}
%\end{widetext}

Equation (19) can be rewritten in full matrix form as

\begin{equation}
%\bigg[i\hbar\frac{\partial}{\partial t}-H_{\rm grp}(t)-\Sigma(t-t^\prime)\bigg]G_{\rm grp}(t,t^\prime)&=&\delta(t-t^\prime)\delta_{nm}\nonumber\\
\bigg[i\hbar\frac{\partial}{\partial t}-H_{\rm grp}(t)-\Sigma_{\rm S}(t-t^\prime)-\Sigma_{\rm D}(t-t^\prime)\bigg]G_{\rm grp}(t,t^\prime)=\delta(t-t^\prime)\delta_{nm}.
\end{equation}

Applying adiabatic expansion on $G_{\rm grp}(t,t^\prime)$ one obtains %\newline

\begin{equation}
%\bigg[i\hbar\frac{\partial}{\partial t}-H_{\rm grp}(t)-\Sigma_{\rm S}(t-t^\prime)-\Sigma_{\rm D}(t-t^\prime)\bigg]G_{\rm grp}\bigg(t-t^\prime,\frac{t+t^\prime}{2}\bigg)&=&\delta(t-t^\prime)\delta_{nm} \nonumber\\
\bigg[i\hbar\frac{\partial}{\partial t}-H_{\rm grp}(t)-\Sigma_{\rm S}(t-t^\prime)-\Sigma_{\rm D}(t-t^\prime)\bigg]\bigg(G^{(0)}_{\rm grp}(t-t^\prime,\bar t)+G^{(1)}_{\rm grp}(t-t^\prime,\bar t)\bigg)=\delta(t-t^\prime)\delta_{nm}.
\end{equation}

Expanding $H_{\rm grp}(t)$ around $\frac{t+t^\prime}{2}=\bar t$ \cite{55} gives rise to
%\begin{widetext}
\begin{equation}
H_{\rm grp}(t)=H_{\rm grp}(\bar t)+\frac {\partial H_{\rm grp}(\bar t)}{\partial \bar t}(t-\bar t)=H^{(0)}_{\rm grp}(\bar t)+H^{(1)}_{\rm grp}(\bar t).
\end{equation}
%\end{widetext}

Substituting Eq. (22) in Eq. (21) to get

\begin{equation}
\bigg[i\hbar\frac{\partial}{\partial t}-H^{(0)}_{\rm grp}(\bar t)-H^{(1)}_{\rm grp}(\bar t)-\Sigma_{\rm S}(t-t^\prime)-\Sigma_{\rm D}(t-t^\prime)\bigg]\bigg(G^{(0)}_{\rm grp}(t-t^\prime,\bar t)+G^{(1)}_{\rm grp}(t-t^\prime,\bar t)\bigg)=\delta(t-t^\prime)\delta_{nm}.
\end{equation}

Consider only the zeroth order contribution gives

\begin{equation}
%\lefteqn{
\bigg[i\hbar\frac{\partial}{\partial t}-H^{(0)}_{\rm grp}(\bar t)-\Sigma_{\rm S}(t-t^\prime)-\Sigma_{\rm D}(t-t^\prime)\bigg]G^{(0)}_{\rm grp}(t-t^\prime,\bar t)=\delta(t-t^\prime)\delta_{nm}.
\end{equation}

Taking the Fourier transform with respect to the fast time variable $(t-t^\prime)$
we get
\begin{equation}
\bigg[\varepsilon-H^{(0)}_{\rm grp}(\bar t)-\Sigma_{\rm S}(\varepsilon)-\Sigma_{\rm D}(\varepsilon)\bigg]G^{(0)}_{\rm grp}(\varepsilon,\bar t)=1.
\end{equation}

This leads to the zeroth order Green's function for graphene in the presence of
the coupling with the contacts
\begin{equation}
G_{\rm grp}^{(0)}(\varepsilon,\bar t)=\frac{1}{\varepsilon-H^{(0)}_{\rm grp}(\bar t)-\Sigma_{\rm S}(\varepsilon)-\Sigma_{\rm D}(\varepsilon)}.
\end{equation}

\subsection{Dyson equation}

Equation (19) in full matrix form is expressed as

\begin{equation}
\bigg[i\hbar\frac{\partial}{\partial t}-\varepsilon_n(t)-\sum_{<nn^\prime>}\gamma_{nn^\prime}(t)\bigg]G_{\rm grp}(t,t^\prime)=\delta (t-t^\prime)\delta_{nm}+ \Sigma(t-t^\prime)G_{\rm grp}(t,t^\prime).
\end{equation}

%%\begin{widetext}
%\begin{eqnarray}
%\lefteqn{\bigg[i\hbar\frac{\partial}{\partial t}-\varepsilon_n(t)-\sum_{<nn^\prime>}\gamma_{nn^\prime}(t)\bigg]G_{\rm grp}(t,t^\prime)}\nonumber\\
%=&&\delta (t-t^\prime)\delta_{nm}+ \Sigma(t-t^\prime)G_{\rm grp}(t,t^\prime).
%\end{eqnarray}
%%\end{widetext}

Define two auxillary time-ordered Green's functions $\rm g$ and $\rm \bar g$ that
satisfy the equation of motions
\begin{equation}
\bigg[i\hbar\frac{\partial}{\partial t}-\varepsilon_n(t)\bigg]{\rm g}(t,t^{\prime})=\delta(t-t^{\prime}),
\end{equation}

and
\begin{equation}
\bigg[i\hbar\frac{\partial}{\partial t}-\varepsilon_n(t)-\sum_{<nn^\prime>}\gamma_{nn^\prime}(t)\bigg]{\rm \bar g}(t,t^\prime)=\delta(t-t^\prime), %\nonumber\\
\end{equation}

which is rearranged to
%\begin{widetext}
\begin{equation}
\bigg[i\hbar\frac{\partial}{\partial t}-\varepsilon_n(t)\bigg]{\rm \bar g}(t,t^\prime)=\delta(t-t^\prime)+\sum \gamma _{nn^\prime}(t){\rm \bar g}(t,t^\prime).
\end{equation}
%\end{widetext}

Using Eq. (28), Eq. (30) can be further written as
\begin{equation}
{\rm \bar g}(t,t^\prime)={\rm g}(t,t^\prime)\bigg[\delta(t-t^\prime)+\sum \gamma _{nn^\prime}(t){\rm \bar g}(t,t^\prime)\bigg].
\end{equation}

From Eqs. (27) and (29)
\begin{equation}
G_{\rm grp}(t,t^\prime)={\rm \bar g}(t,t^\prime)+{\rm \bar g}(t,t^\prime)\Sigma(t-t^\prime)G_{\rm grp}(t,t^\prime),
\end{equation}

which is the Dyson equation for the system of gas sensor.

\section{Results and Discussions}
\label{4}
\subsection{Zeroth order time-dependent Green's function}
The zeroth order graphene Hamiltonian $H^{(0)}_{\rm grp}(\bar t)$ is calculated by
expanding the Hamiltonian $H_{\rm grp}(t)$ given in Eq. (6) around $\bar t$ using Eq.
(22), and taking only the zeroth order contribution. Using Eq. (26) and the Hamiltonian
$H^{(0)}_{\rm grp}(\bar t)$, the zeroth order time-dependent retarded Green's function
for graphene $G^{(0)r}_{\rm grp}$ is calculated for a 9 $\times$ 9 matrix and is given by

\begin{equation}
%\lefteqn{
G^{(0) r}_{\rm grp}(\varepsilon,\bar{t})=\\
\left(
\begin{array}{ccccccccc}
\varepsilon-\varepsilon_{\rm A_{1}}(\bar{t})-\Sigma^r_{\rm S}(\varepsilon) & {\gamma_{11}(\bar{t})} &  &  &&\\
{\gamma_{11}(\bar{t})} & \varepsilon-{\varepsilon_{\rm B_{1}}(\bar{t})} & {\gamma_{12}(\bar{t})} & &&\\
& {\gamma_{12}(\bar{t})} & \varepsilon-{\varepsilon_{\rm A_{2}}(\bar{t})} & \ddots&& \\
&& \ddots & \ddots &\ddots &\\
&&&\ddots&\varepsilon-\varepsilon_{\rm B_4}(\bar{t})&\gamma_{45}(\bar{t})\\
&&&&\gamma_{45}(\bar{t})&\varepsilon-\varepsilon_{\rm A_5}(\bar{t})-\Sigma^r_{\rm D}(\varepsilon)
\end{array} \right)_{9\times 9}^{-1}.
\end{equation}

Using the Green's function the density of states can be calculated by $D(\varepsilon)=i[G^r-G^a]/2\pi$.
\subsection{Time-dependent Landauer formula}

Using Eq. (15) and assuming the source and drain coupling functions are proportional
to each other $({\Gamma}_{\rm S}(\varepsilon)=\lambda{\Gamma}_{\rm D}(\varepsilon))$,
and writing the current as $I^{(0)}(\bar t)=x I_{\rm S}^{(0)}-(1-x)I_{\rm D}^{(0)}$
\cite{51} the expression for the total current is expressed as

\begin{eqnarray}
I^{(0)}(\bar t)&=&\frac{ie}{\hbar}\int\frac{d\varepsilon}{2\pi}{\rm Tr}\bigg\{\bigg[x\lambda -1+x\bigg]\Gamma_{\rm D}(\varepsilon)
G^{(0)<}_{\rm grp}(\varepsilon, \bar t)+\bigg[x\lambda f_{\rm S}(\varepsilon)-f_{\rm D}(\varepsilon)+xf_{\rm D}(\varepsilon)\bigg]\nonumber\\
&&\times\Gamma_{\rm D}(\varepsilon)\bigg[{G}^{(0)r}_{\rm grp}(\varepsilon, \bar t)-{G}^{(0)a}_{\rm grp}(\varepsilon, \bar t)\bigg]\bigg\}.
\end{eqnarray}

Now fixing the arbitrary parameter $x=\frac{1}{(1+\lambda)}$, where $\lambda$ is the
constant of proportionality, a simple expression for the total current through graphene
is derived as

\begin{equation}
I^{(0)}(\bar t )=\frac{ie}{\hbar}\int\frac{d\varepsilon}{2\pi}{\rm Tr}\bigg\{\frac{\Gamma_{\rm S}(\varepsilon)\Gamma_{\rm D}(\varepsilon)}{\Gamma_{\rm S}(\varepsilon)+\Gamma_{\rm D}(\varepsilon)}\bigg({G}^{(0)r}_{\rm grp}(\varepsilon,
\bar t)-{G}^{(0)a}_{\rm grp}(\varepsilon, \bar t)\bigg)\bigg\}[f_{\rm S}(\varepsilon)-f_{\rm D}(\varepsilon)],
\end{equation}

Applying the adiabatic expansion and taking the Fourier transform of the Dyson equation
(Eq. (32)) for the retarded Green's function, one obtains
\begin{equation}
G^r_{\rm grp}(\varepsilon,\bar t)={\rm \bar g}^r(\varepsilon,\bar t)+{\rm \bar g}^r(\varepsilon,\bar t)\Sigma^r(\varepsilon)G^r_{\rm grp}(\varepsilon,\bar t),
\end{equation}

where $\Sigma^r(\varepsilon)$ is the self energy which can be expressed as \cite{56}
\begin{equation}
\Sigma^r(\varepsilon)=\frac{G^{r}_{\rm grp}(\varepsilon,\bar t)-{\rm \bar g}^r(\varepsilon,\bar t)}{{\bar g}^r(\varepsilon,\bar t)G^{r}_{\rm grp}(\varepsilon,\bar t)}={\rm \bar g}^r(\varepsilon,\bar t)^{-1}-G^{r}_{\rm grp}(\varepsilon,\bar t)^{-1}.
\end{equation}

In the present model system, the effect of vapor interaction with the DNA coated graphene
is included in the Hamiltonian $H_{\rm grp}(\bar t)$ (as discussed before in section II)
not in the self-energy which is time-independent and includes only the tunneling contributions
with no interaction. The retarded and advanced self-energies are defined as
\begin{equation}
\Sigma^{r,a}(\varepsilon)=\sum_{k\alpha\epsilon{\rm S, D}}|V_{k\alpha}|^2g^{r,a}_{k\alpha}(\varepsilon)=\Lambda(\varepsilon)\mp i\Gamma(\varepsilon)/2,
\end{equation}
where $\Lambda(\varepsilon)$ and $\Gamma(\varepsilon)/2$ are the real and imaginary parts
of the self energy with $\Lambda(\varepsilon)=\Lambda_{\rm S}(\varepsilon)+\Lambda_{\rm D}
(\varepsilon)$ and $\Gamma(\varepsilon)=\Gamma_{\rm S}(\varepsilon)+\Gamma_{\rm D}(\varepsilon)$
\cite{51}.

Equation (38) leads to the self-energy difference \cite{56} as
\begin{equation}
\Sigma(\varepsilon)=\Sigma^r(\varepsilon)-\Sigma^a(\varepsilon)=-i\Gamma(\varepsilon).
\end{equation}

Using Eq. (38) the zeroth order retarded and advanced Green's functions for graphene from
Eq. (26) are

\begin{equation}
G^{(0)r,a}_{\rm grp}(\varepsilon,\bar t)=[\varepsilon-H^{(0)}_{\rm grp}(\bar t)-\Lambda(\varepsilon)\pm i\Gamma(\varepsilon)/2]^{-1}.
\end{equation}

This equation gives rise to
\begin{equation}
G^{(0)r}_{\rm grp}(\varepsilon,\bar t)-G^{(0)a}_{\rm grp}(\varepsilon,\bar t)=\frac{-i\Gamma(\varepsilon)}{(\varepsilon-H^{(0)}_{\rm grp}(\bar t)-\Lambda(\varepsilon))^2+(\Gamma(\varepsilon)/2)^2},
\end{equation}

and

\begin{equation}
G^{(0)r}_{\rm grp}(\varepsilon,\bar t)G^{(0)a}_{\rm grp}(\varepsilon,\bar t)=\frac{1}{(\varepsilon-H^{(0)}_{\rm grp}(\bar t)-\Lambda(\varepsilon))^2+(\Gamma(\varepsilon)/2)^2}.
\end{equation}

From Eqs. (41) and (42)
\begin{eqnarray}
G^{(0)r}_{\rm grp}(\varepsilon,\bar t)-G^{(0)a}_{\rm grp}(\varepsilon,\bar t)&=&G^{(0)r}_{\rm grp}(\varepsilon,\bar t)G^{(0)a}_{\rm grp}(\varepsilon,\bar t)(-i\Gamma(\varepsilon))\nonumber\\
&=&G^{(0)r}_{\rm grp}(\varepsilon,\bar t)(-i\Gamma(\varepsilon))G^{(0)a}_{\rm grp}(\varepsilon,\bar t).
\end{eqnarray}

Using Eq. (39), Eq.(43) becomes
\begin{eqnarray}
G^{(0)r}_{\rm grp}(\varepsilon,\bar t)-G^{(0)a}_{\rm grp}(\varepsilon,\bar t)&=&G^{(0)r}_{\rm grp}(\varepsilon,\bar t)(-i\Gamma(\varepsilon)) G^{(0)a}_{\rm grp}(\varepsilon,\bar t)\nonumber\\
&=&G^{(0)r}_{\rm grp}(\varepsilon,\bar t)\Sigma(\varepsilon) G^{(0)a}_{\rm grp}(\varepsilon,\bar t).
\end{eqnarray}

Substituting Eq. (44) in Eq. (35)
\begin{eqnarray*}
I^{(0)}(\bar t)&=&
%\frac{ie}{\hbar}\int\frac{d\varepsilon}{2\pi} {\rm Tr} \bigg\{\Gamma(\varepsilon){G}^{(0)r}_{\rm grp}(\varepsilon, \bar t)\Sigma(\varepsilon){G}^{(0)a}_{\rm grp}(\varepsilon, \bar t)\bigg\}[f_{\rm S}(\varepsilon)-f_{\rm D}(\varepsilon)]\nonumber\\
%&=&
\frac{e}{\hbar}\int\frac{d\varepsilon}{2\pi} {\rm Tr} \bigg\{\frac{i\Gamma_{\rm S}(\varepsilon)\Gamma_{\rm D}(\varepsilon)}{\Gamma_{\rm S}(\varepsilon)+\Gamma_{\rm D}(\varepsilon)}{G}^{(0)r}_{\rm grp}(\varepsilon, \bar t)\Sigma(\varepsilon){G}^{(0)a}_{\rm grp}(\varepsilon, \bar t)\bigg\}[f_{\rm S}(\varepsilon)-f_{\rm D}(\varepsilon)]. %\nonumber\\
%&=&\frac{e}{\hbar}\int\frac{d\varepsilon}{2\pi} {\rm Tr} \bigg\{\Gamma_{\rm S}(\varepsilon)\Gamma_{\rm D}(\varepsilon){G}^{(0)r}_{\rm grp}(\varepsilon, \bar t)\Sigma(\varepsilon)\Sigma^{-1}_0(\varepsilon){G}^{(0)a}_{\rm grp}(\varepsilon, \bar t)\bigg\}[f_{\rm S}(\varepsilon)-f_{\rm D}(\varepsilon)]\nonumber\\
%&=& \frac{e}{\hbar}\int\frac{d\varepsilon}{2\pi} {\rm Tr}\bigg\{\Gamma_{\rm S}(\varepsilon){G}^{(0)r}_{\rm grp}(\varepsilon, \bar t)\Gamma^{\rm D}(\varepsilon){G}^{(0)a}_{\rm grp}(\varepsilon, \bar t)\Sigma(\varepsilon)\Sigma^{-1}_0(\varepsilon)\bigg\}[f_{\rm S}(\varepsilon)-f_{\rm D}(\varepsilon)],
\end{eqnarray*}
In this equation, the quantity $-i(\Gamma_{\rm S}(\varepsilon)+\Gamma_{\rm D}(\varepsilon))=-i\Gamma(\varepsilon)$
is just the self-energy $\Sigma(\varepsilon)$ (Eq. (39)). Hence the current becomes
\begin{equation}
I^{(0)}(\bar t)
%&=&\frac{e}{\hbar}\int\frac{d\varepsilon}{2\pi} {\rm Tr} \bigg\{\Gamma_{\rm S}(\varepsilon)\Gamma_{\rm D}(\varepsilon){G}^{(0)r}_{\rm grp}(\varepsilon, \bar t)\Sigma(\varepsilon)\Sigma^{-1}(\varepsilon){G}^{(0)a}_{\rm grp}(\varepsilon, \bar t)\bigg\}[f_{\rm S}(\varepsilon)-f_{\rm D}(\varepsilon)]\nonumber\\
= \frac{e}{\hbar}\int\frac{d\varepsilon}{2\pi} {\rm Tr}\bigg\{\Gamma_{\rm S}(\varepsilon){G}^{(0)r}_{\rm grp}(\varepsilon, \bar t)\Gamma_{\rm D}(\varepsilon){G}^{(0)a}_{\rm grp}(\varepsilon, \bar t)\Sigma(\varepsilon)\Sigma^{-1}(\varepsilon)\bigg\}[f_{\rm S}(\varepsilon)-f_{\rm D}(\varepsilon)],
\end{equation}
which leads to
%here $-i(\Gamma^{\rm S}(\varepsilon)+\Gamma^{\rm D}(\varepsilon))=\Sigma_0(\varepsilon)$ which is the self-energy for the noninteracting case [18].
%Since the interactions are included in $G(\varepsilon,\bar t)$ therefore $\Sigma(\varepsilon)\sim\Sigma_0(\varepsilon)$,
%resulting in
\begin{equation}
I^{(0)}(\bar t)=\frac{e}{\hbar}\int\frac{d\varepsilon}{2\pi} {\rm Tr} \bigg\{\Gamma_{\rm S}(\varepsilon){G}_{\rm grp}^{(0)r}(\varepsilon, \bar t)\Gamma_{\rm D}(\varepsilon){G}_{\rm grp}^{(0)a}(\varepsilon, \bar t)\bigg\}[f_{\rm S}(\varepsilon)-f_{\rm D}(\varepsilon)].
\end{equation}

This is the time-dependent Landauer formula derived in terms of the slow time variable
$\bar t$, where $f_{\{\rm S, D\}}(\varepsilon)$ are the Fermi distribution functions
in the source and drain contacts, and $\Gamma_{\{\rm S, D\}}=i[\Sigma^r_{\{\rm S, D\}
}-\Sigma^a_{\{\rm S, D\}}]$ describe the contact-graphene coupling.

\subsection{The transmission function}
From equation (46) the transmission function of the graphene system is identified as
\begin{equation}
T(\varepsilon,\bar{t})={\rm Tr}({\Gamma_{\rm S}}(\varepsilon){G^{(0){r}}_{\rm grp}
(\varepsilon,\bar{t})}{\Gamma_{\rm D}}(\varepsilon){G^{(0){a}}_{\rm grp}
(\varepsilon,\bar{t})}),
\end{equation}

Considering the first element of the $\Gamma_{\rm S}(\varepsilon)(\Sigma_{\rm S}(\varepsilon))$
matrix and the last element of the $\Gamma_{\rm D}(\varepsilon)(\Sigma_{\rm D}(\varepsilon))$
matrix the transmission function can be rewritten as
\begin{equation}
T(\varepsilon,\bar{t})={\Gamma_{\rm S,11}(\varepsilon)}{G^{(0)r}_{\rm grp19}(\varepsilon,\bar{t})}
{\Gamma_{\rm D,99}(\varepsilon)}{G^{(0)r*}_{\rm grp19}(\varepsilon,\bar{t})},
\end{equation}

where $G^{(0)r*}={G^{(0)a}}$.

To calculate the current explicitly, an expression for the transmission function is
derived in the linear response \cite{50} regime as the experiment is performed with
low bias. In linear response, the expression for the current (Eq. (46)) becomes
\begin{equation}
I^{(0)}(\bar{t})\approx\frac{e}{\hbar}\int{\frac{\rm d {\varepsilon}}{2\pi}}
T({\varepsilon},\bar{t})\delta[f_{\rm S}(\varepsilon)-f_{\rm D}(\varepsilon)].
\end{equation}

Equation (49) can be further written as
\begin{equation}
I^{(0)}(\bar{t})\sim\frac{e}{\hbar}T({\varepsilon_f},\bar{t})[\mu_{\rm S}-\mu_{\rm D}],
\end{equation}

where $\mu_{\{\rm S,\rm D\}}$ are the chemical potentials associated with the S and D
contacts \cite{50}. Equation (50) is obtained by using $\delta[f_{\rm S}(\varepsilon)-f_{\rm D}
(\varepsilon)]=(\mu_{\rm S}-\mu_{\rm D})(-\frac{\partial f_0}{\partial\varepsilon})$
and $(-\frac{\partial f_0}{\partial\varepsilon})=\delta(\varepsilon_f-\varepsilon)$
\cite{50} where $f_0(\varepsilon)=1/[1+{\rm exp}((\varepsilon-\mu)/k_BT)]$ is the
equilibrium Fermi function and $\varepsilon_f$ is the Fermi energy of graphene.

Transport is often dominated by states close to the Fermi level, and $\Gamma(\varepsilon)$
and $\Lambda(\varepsilon)$ the imaginary and real parts of the self-energy are generally
slowly varying functions of energy, therefore the wide-band limit is considered in which
the real part ($\Lambda$) of the self energy is neglected and the imaginary part ($\Gamma$)
is considered to be energy independent. This approximation has the advantage of providing
explicit analytic results \cite{54,55}. Hence, $\Lambda$ is neglected and $\Gamma$ is
considered energy independent from now onwards. Taking $\Gamma_{\rm S,11}= -2 {\rm Im}
(\Sigma^r_{\rm S,11})=-2 {\rm Im}\Sigma^r_{\rm S}$ and $\Gamma_{\rm D,99}= -2 {\rm Im}
(\Sigma^r_{\rm D,99})= -2 {\rm Im}\Sigma^r_{\rm D}$ from the relationship $\Gamma_{\{\rm
S, D\}}=i[\Sigma^r_{\{\rm S, D\}}-\Sigma^a{_{\{\rm S, D\}}}]$, and using Eqs. (33) and
(48) the expression for the transmission function for graphene coated with ssDNA sequence
1 in linear response regime is given by

\begin{equation}
%\fl
T(\varepsilon_f,\bar{t})=\frac{4~{\rm Im}{\Sigma^r_{\rm S}}~{\rm Im}{\Sigma^r_{\rm D}}~{\gamma^{2}_{11}(\bar{t})}
{\gamma^{2}_{12}(\bar{t})}{\gamma^{2}_{22}(\bar{t})}{\gamma^{2}_{23}(\bar{t})}{\gamma^{2}_{33}(\bar{t})}
{\gamma^{2}_{34}(\bar{t})}{\gamma^{2}_{44}(\bar{t})}{\gamma^{2}_{{45}}}(\bar{t})}{|{G^{(0)r}_{\rm grp}}
(\varepsilon_f,\bar{t})|^2_{9\times9}}.
\end{equation}

\subsection{The first order contribution to the current}

The first order contribution to the current is expressed as

\begin{equation}
I^{(1)}(\bar t )=\frac{ie}{\hbar}\int\frac{d\varepsilon}{2\pi}{\rm Tr}\bigg\{\frac{\Gamma_{\rm S}(\varepsilon)\Gamma_{\rm D}
(\varepsilon)}{\Gamma_{\rm S}(\varepsilon)+\Gamma_{\rm D}(\varepsilon)}\bigg({G}^{(1)r}_{\rm grp}(\varepsilon,
\bar t)-{G}^{(1)a}_{\rm grp}(\varepsilon, \bar t)\bigg)\bigg\}[f_{\rm S}(\varepsilon)-f_{\rm D}(\varepsilon)].
\end{equation}

The first order current $I^{(1)}(\bar t)$ is found small as compared to the zeroth
order current $I^{(0)}(\bar{t})$. The first order Green's function $G^{(1)}(\bar t)$
gives rise to the first order current $I^{(1)}(\bar t)$, and it is defined as
$G^{(1)r,a}(t-t^{\prime},\bar{t})=(\frac{t^{\prime}-t}{2})\frac{\partial G^{r,a}}
{\partial\bar{t}}(t-t^\prime,\bar{t})|_{\bar{t}=t}$ where the fast time variable
$(t-t^\prime)$ is of the order of $\sim 10^{-15}$ s. Therefore, the first order
contribution being very small is not included in the final expression for the
total current. There could be future experiments in which the first order term
is enhanced (\emph{e.g.,} when $\frac{\partial G^{r,a}}{\partial \bar t}\sim
\frac{1}{t^\prime-t}$) and the first order contribution becomes significant.

\section{Theory and experiment}
\label{5}
To show that the theory and experiment are in a good agreement the sensor response,
$\Delta I^{(0)}(\bar t)/I_0$ $=1-I^{(0)}(\bar t)({\rm vapor})/I_0({\rm air})=1-T
(\varepsilon_f,\bar t)/T_0(\varepsilon_f^\prime,\bar t_{\rm max})$ of the DNA
coated graphene device is explicitly calculated using a $9\times9$ matrix and a
form for the hopping integral given as $\gamma_{\rm ij}(\Delta {\rm a_{cc}}(\bar t))
=\gamma_0 {\rm exp}(-\Delta {\rm a_{cc}}(\bar t)/{\rm a_0})$ \cite{57}. This form for the
hopping integral is derived from the definition of the hopping integral which is
given as $\gamma=<\varphi_{\rm A(r-R_A)}|H_{\rm grp}|\varphi_{\rm B(r-R_B)}>$,
where $\varphi_{\rm A}$ and $\varphi_{\rm B}$ are the atomic wavefunctions of the
carbon atoms A and B, and $R_{\rm A(B)}$ are their position vectors. Using the
standard wavefunctions, this form for the hopping integral can be found. % to have the form $\gamma_{\rm ij}
%(\Delta {\rm a_{cc}}(\bar{t}))= \gamma_0 {\rm exp(-\Delta a_{cc}}(\bar{t})/{\rm a_0})$.
The parameter $\gamma_0$ is the hopping integral of the DNA coated graphene in the
absence of DMMP vapor (only for $\rm N_2$) which is chosen to be 3.3 eV with $\rm
a_0=0.33 \rm \AA$. The value of the hopping integral $\gamma_{\rm ij}(\Delta {\rm
a_{cc}}(\bar{t}))$ changes when the nearest-neighbor carbon-carbon distance $\rm
a_{cc}$ changes by $\Delta \rm a_{cc}$ as a function of time $\bar{t}$ due to the
interaction of vapor molecules at each time $\bar{t}$ s with the DNA-decorated graphene.
The $9 \times 9$ matrix consists of nine carbon atoms that are involved in the transport.
Out of these nine carbon atoms the first $(\rm A_1)$ and the last $(\rm X_{(M+1)/2})$
carbon atoms are connected to the source and drain electrodes. Seven ssDNA bases G,
A, G, T, C, T, and G  of sequence 1 interact with seven carbon atoms excepting the
first and the last that are connected to S and D electrodes, Fig. 1. This is how the
complex Gas-DNA-graphene system is modeled in a simplified way to explicitly calculate
the sensor response.

The transmission function $T(\varepsilon_f,\bar t)$ of the DNA-decorated graphene
when exposed to DMMP vapor is calculated at different times $\bar t$ with varying
Fermi level $\varepsilon_f$, whereas  $T_0(\varepsilon_f^\prime,\bar t_{\rm max})=
\frac{4~{\rm Im}{\Sigma^r_{\rm S}}~{\rm Im}{\Sigma^r_{\rm D}}~{\gamma^{2}_{0}}
{\gamma^{2}_{0}}{\gamma^{2}_{0}}{\gamma^{2}_{0}}{\gamma^{2}_{0}}{\gamma^{2}_{0}}
{\gamma^{2}_{0}}{\gamma^{2}_{0}}}{|{G^{(0)r}_{\rm grp}}(\varepsilon_f^\prime, \bar
t_{\rm max})|^2_{9 \times 9}}$ is the transmission function of the device when exposed
to $\rm N_2$ and calculated at the maximum time when all $\rm N_2$ molecules interact
with the full DNA sequence, with a fixed value of the Fermi level $\varepsilon_f^\prime$.
The best fitted value of $\varepsilon_f^\prime$ is found to be -0.7 eV as in this
case the devices is hole doped due to $\rm N_2$ and the Fermi level lies in the valence
band.

To reproduce the experimental results, the model parameters $\Delta \rm a_{cc}$ and
$\varepsilon_f$ are fitted, and the best fitted values are given in Table I. These
model parameters are found to be specific to the bases of the DNA sequence. The
values of ${\rm a_{cc}}(\bar t)$ and $\varepsilon_f(\bar t)$ are found to decrease
(becomes more negative) and increase, respectively with time $\bar t$ leading to a
decrease in sensor current. In the graphene channel, hole conduction dominates
\cite{37} and DMMP is a strong electron donor \cite{58}. On interaction of DMMP
vapor molecules, at a particular concentration, with ssDNA coated graphene a
fractional charge transfer takes place from each DMMP-DNA-base complex to the
graphene channel. This causes deformation in the carbon lattice of graphene and
decreases the nearest-neighbor carbon-carbon distance $\rm a_{cc}$ from its values
when there is no vapor which affects the wavefunctions and enhances the hopping
integral. As a result, the energy bands of graphene shift downward to lower energy
which moves the Dirac point downward towards a lower energy position, resulting in
a shift in the position of the Fermi level $(E_f)$ via the relationship $\Delta
E_f=E_f-E_{\rm Dirac}$. %[]
%As a result, the energy bands of graphene change which shifts the position of the Fermi level. The
%increasing values of $\varepsilon_f(\bar t)$ in Table 1 indicates that the Fermi level moves away from
%the valence band of graphene, reducing the hole carrier concentration and hence the device current, which
%is consistent with the experiment.

Referring to Fig. 1(a) and Table I, it is observed that at time $\bar t_1$ when
the DMMP interacts with guanine base the change in C-C is ${\rm \Delta a_{\rm A_1B_1}}
(\bar t_1)=-0.02$ and ${\rm \Delta a_{\rm B_1A_2}}(\bar t_1)=-0.018$. At time $\bar t_2$
DMMP interacts with another base adenine which causes a change ${\rm \Delta a_{B_1A_2}}
(\bar t_2)=-0.018-0.011=-0.029$ and ${\rm \Delta a_{A_2B_2}}(\bar t_2) =-0.009$, while
${\rm \Delta a_{\rm A_1B_1}}(\bar t_2)$ remain unchanged. At time $\bar t_3$ DMMP
interacts with guanine again which causes a change ${\rm \Delta a_{A_2B_2}}(\bar t_3)
=-0.009-0.02=-0.029$ and ${\rm \Delta a_{B_2A_3}}(\bar t_3)=-0.018$. At time $\bar t_4$
DMMP interacts with thymine giving change ${\rm \Delta a_{B_2A_3}}(\bar t_4)=-0.018-0.012
=-0.03$ and ${\rm \Delta a_{A_3B_3}}(\bar t_4)=-0.01$. At time $\bar t_5$ DMMP interacts
with cytosine and the change is ${\rm \Delta a_{A_3B_3}}(\bar t_5)=-0.01-0.009=-0.019$
and ${\rm \Delta a_{B_3A_4}}(\bar t_5)=-0.007$. At time $\bar t_6$ DMMP interacts with
thymine and the change is ${\rm \Delta a_{B_3A_4}}(\bar t_6)=-0.007-0.012=-0.019$ and
${\rm \Delta a_{A_4B_4}}(\bar t_6)=-0.01$. At time $\bar t_7$ DMMP interacts with
guanine and causing a change ${\rm \Delta a_{A_4B_4}}(\bar t_7)=-0.01-0.02=-0.03$ and
${\rm \Delta a_{B_4A_5}}(\bar t_7)=-0.018$. Hence, the change in nearest-neighbor C-C
distance due to interaction of DMMP with different bases, which changes the sensor
response, is studied explicitly. Figure 2 shows a good match of the calculated sensor
response and the experimentally measured sensor response of ssDNA sequence 1 coated
graphene device as a function of the exposure time of DMMP vapor. When the device
is exposed to pure $\rm N_2$ gas the response is found in the opposite direction. In
this case, the charge transfer takes place from graphene to the $\rm N_2$-DNA-base
complex causing an increase in ${\rm a_{cc}}(\bar t)$ from its modified value due to
vapor and a decrease in $\varepsilon_f(\bar t)$. This results in lowering the hopping
integral and shifting the Fermi level closer to the valence band as shown by decreasing
values of $\varepsilon_f(\bar t)$ in Table II. This increases the hole carrier concentration
hence the device current which is in agreement with the experiment. Figure 3 shows the
schematic of the interaction of $\rm N_2$ with DNA coated graphene. When $\rm N_2$
interacts with guanine at time $\bar t_1$ the change in $\rm a_{cc}$ is ${\rm \Delta
a_{A_1B_1}}(\bar{t_1})=-0.02+0.04=0.02$ and  ${\rm \Delta a_{B_1A_2}}(\bar{t_1})=-0.029
+0.02=-0.009$. At time $\bar t_2$ the change in $\rm a_{cc}$ due to adenine is ${\rm
\Delta a_{B_1A_2}}(\bar{t_2})=-0.009+0.013=0.004$ and ${\rm \Delta a_{A_2B_2}}(\bar{t_2})
=-0.029+0.011=-0.018$. In a similar manner, the change in $\rm a_{cc}$ due to guanine
base at $\bar t_3$ is ${\rm \Delta a_{A_2B_2}}(\bar{t_3})=-0.018+0.04=0.022$ and
${\rm \Delta a_{B_2A_3}}(\bar{t_3})=-0.03+0.02=-0.01$. At time $\bar t_4$, the value
of $\rm \Delta a_{cc}$ due to thymine at $\bar t_4$ is  ${\rm \Delta a_{B_2A_3}}
(\bar{t_4})=-0.01+0.014=0.004$ and ${\rm \Delta a_{A_3B_3}}(\bar{t_4})=-0.019+0.012
=-\rm 0.007$. At $\bar t_5$ the change in $\rm a_{cc}$ due to cytosine is ${\rm
\Delta a_{A_3B_3}}(\bar{t_5})=-0.007+0.011=0.004$ and ${\rm \Delta a_{B_3A_4}}(\bar{t_5})
=-0.019+0.009=-0.01$. At $\bar t_6$ the change in $\rm a_{cc}$ due to thymine is
${\rm \Delta a_{B_3A_4}}(\bar {t_6})=-0.01+0.014=0.004$ and ${\rm \Delta a_{A_4B_4}}
(\bar t_6)=-0.03+0.012=-0.018$. At $\bar t_7$ the change in $\rm a_{cc}$ due to
guanine is ${\rm \Delta a_{A_4B_4}}(\bar t_7)=-0.018+0.04=0.022$ and ${\rm \Delta
a_{B_4A_5}}(\bar t_7)=-0.018+0.02=0.002$. Figure 4 compares the theoretical and
experimental results of the device with respect to exposure time of $\rm N_2$.
The values for different parameters used in the calculation of the current for
$\rm N_2$ are given in Table II. Detailed steps of the calculation of the sensor
response are given in Appendix A. Similar calculations are also done for the
graphene device when coated with ssDNA sequence 2 and exposed to DMMP and
$\rm N_2$. The values of the fitted parameters are given in Tables III and IV,
and the sensor responses are shown in Figs. 5 and 6.  Hence, the formula derived
for the transmission function reproduces the sensor response of the experiment.
The calculations show that the change in $\rm a_{cc}$ due to interaction of vapor
with four bases G, A, T and C is in the order of $\rm G>T\sim A>C$ indicating the
order of interaction energy of these bases towards graphene. The order of the
interaction strength of these bases with graphene is also found in other theoretical
and experimental studies where the predicted orders are different, e.g.,
$(\rm G>A\approx T\approx C)$, $(\rm G>A\sim T>C)$, $(\rm G>A>C<T)$ or $\rm (G>A>C>T)$
\cite{59,60} and is not conclusive. These results are the prediction of the model
which should be verified by experiments, ab-initio technique or density functional
theory based calculations.
%More sophisticated methods must be tried on these systems
%for better modeling the DNA and this is left for future studies.

Experimental result (Fig. 2 of Ref. [37]) shows that the current response of devices
to DMMP changes (decreases) drastically at progressively larger concentrations. The
model explains the reason for this change as the graphene device is functionalized
with many strands of the same ssDNA sequence 1 or 2 more DNA bases are available for
DMMP molecules to interact with. As the concentration increases the number of vapor
molecules increases and they interact with more DNA strands (bases) causing more
charge transfer to graphene. This results in more changes in C-C distance ($\rm a_{cc}$
becomes more negative) with time which further moves the Dirac point downwards and
shifts the Fermi level further away from the valence band of graphene. Hence reducing
the carrier concentration and the current of the graphene device drastically.

\section{Conclusions}
\label{6}
A quantitative understanding of the experiment based on the DNA functionalized
graphene gas sensor was presented. To study the electronic transport in the gas
sensor a theoretical model of the DNA functionalized graphene was built, and a
connection between the theoretical predictions and experimental observations was
developed.
%A quantitative understanding of the experiment based on the DNA functionalized graphene gas sensor was
%presented which involved building up a theoretical model of the DNA functionalized graphene sensor to
%study electronic transport. A procedure for the development of a connection between the theoretical predictions and
%experimental observations was also outlined.

A time-dependent electronic transport through the graphene device was investigated
based on a tight-binding model of the experimental system. The analytical
calculations of electronic transport were performed using the method of the
NEGF formalism and the adiabatic approximation. The detailed work included
derivations of the Green's function for graphene, and the expression for the
time-dependent current by deriving the equation of motion and Dyson equation.
The zeroth and the first order contributions to the current were derived.
The contribution of the zeroth order current was significant and was identified
as the Landauer formula which is time-dependent in terms of the slow time variable,
whereas the first order contribution was found to be small and hence was not
included in the total current when comparing with experiment. A transmission
function formula was derived in terms of the time-dependent hopping integral
and on-site energy. The sensor responses (change in current) of graphene coated
with ssDNA sequence 1 and 2 for DMMP vapor and $\rm N_2$ were then explicitly
calculated using the transmission function formula and considering a form
for the hopping integral which accounts for the effect of the DNA and DMMP
molecules on the charge distribution of the carbon atoms and the nearest-neighbor
carbon-carbon distance. The sensor response was found to be sensitive to the DNA
bases as the effect of interaction of vapor molecules with individual DNA base
molecules, given by changes in $\rm a_{cc}$, on the electronic transport of
graphene was studied quantitatively. For DMMP and $\rm N_2$, values of $\rm
\Delta a_{cc}$ and $\varepsilon_f$ were found which were used to calculate the
theoretical sensor response for two different DNA sequences and to compare with
the experiment. The calculation suggests that for a particular target vapor,
the same values of $\rm \Delta a_{cc}$ due to the four DNA bases can predict the theoretical
values of the current response for different DNA sequences. These calculations reproduced
the experimental response and demonstrated a significant step towards an
understanding of how the functionalization of graphene with DNA affects its electronic
transport properties when used for detection of chemical vapors and hints on its use as a
DNA analyser.

\begin{table*}
\caption{\label{tabone}Values for different model parameters used in the calculation of the current
with $\gamma_0$=3.3eV and $\varepsilon^\prime_f$=-0.7 eV for DMMP with DNA sequence 1.}
\label{log}
\begin{center}
\begin{tabular}{llllllll} \hline\hline \noalign{\smallskip}
$\bar{t}(s)$&$\bar{t_1}$&$\bar{t_2}$&$\bar{t_3}$&$\bar{t_4}$&$\bar{t_5}$&$\bar{t_6}$&$\bar{t_7}$\\
\hline
$\Delta I^{(0)}(\bar{t})/{I_0}$&$-0.3084$&$-1.1369$&$-1.5444$&$-1.7050$&$-1.9041$&$-2.0848$&$-2.3088$\\
$\varepsilon_f(\bar{t})(eV)$~&1.53~&1.67~&1.75~&1.76~&1.79~&1.81~&1.84\\
${\rm \Delta a_{A_1B_1}}(\bar{t})$(\AA)&$-0.02 $&$-0.02 $&$-0.02 $&$-0.02 $&$-0.02 $&$-0.02 $&$-0.02$\\
${\rm \Delta a_{B_1A_2}}(\bar{t})$(\AA)&$-0.018$&$-0.029$&$-0.029$&$-0.029$&$-0.029$&$-0.029$&$-0.029$\\
${\rm \Delta a_{A_2B_2}}(\bar{t})$(\AA)&---     &$-0.009$&$-0.029$&$-0.029$&$-0.029$&$-0.029$&$-0.029$\\
${\rm \Delta a_{B_2A_3}}(\bar{t})$(\AA)&--- &---         &$-0.018$&$-0.03 $&$-0.03 $&$-0.03 $&$-0.03$\\
${\rm \Delta a_{A_3B_3}}(\bar{t})$(\AA)&--- &--- &---             &$-0.01 $&$-0.019$&$-0.019$&$-0.019$\\
${\rm \Delta a_{B_3A_4}}(\bar{t})$(\AA)&--- &--- &--- &---                 &$-0.007$&$-0.019$&$-0.019$\\
${\rm \Delta a_{A_4B_4}}(\bar{t})$(\AA)&--- &--- &--- &--- &---                     &$-0.01 $&$-0.03$\\
${\rm \Delta a_{B_4A_5}}(\bar{t})$(\AA)&--- &--- &--- &--- &--- &---                         &$-0.018$\\
\noalign{\smallskip} \hline\hline
\noalign{\smallskip}
\end{tabular}
\end{center}
\end{table*}

\begin{figure}[h!]
\center
\includegraphics[width=10cm]{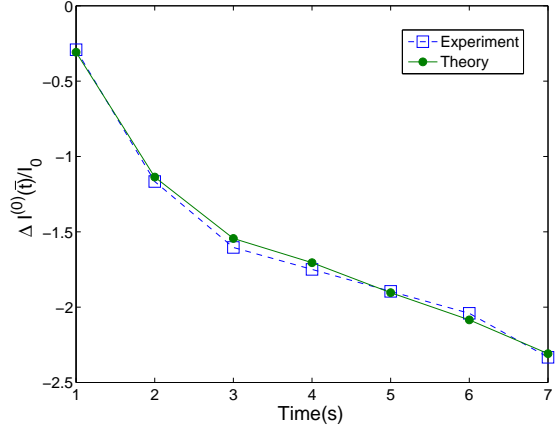}
\caption{The sensor response $\Delta I^{(0)}(\bar{t})/{I_0}$ versus exposure time $\bar{t}$(s)
for DMMP with DNA sequence 1. This plot shows agreement between the theory and experiment.
Experimental data is used from Ref. [37].}
\label{fig.2}
\end{figure}

\begin{figure}[h!]
\center
\includegraphics[width=10cm]{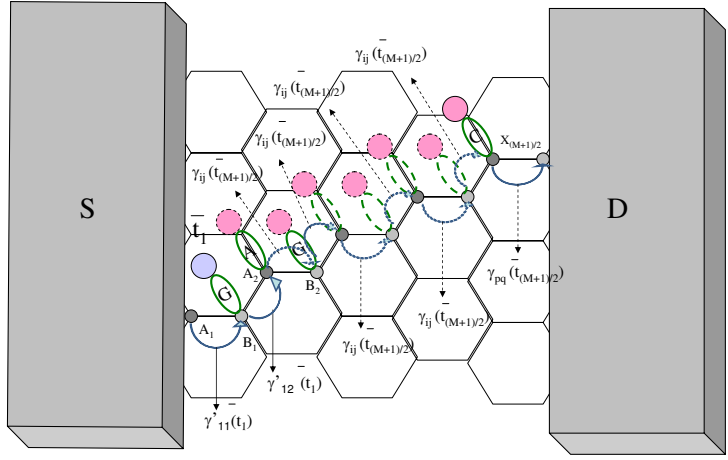}
\caption{Schematic of the interaction of $\rm N_2$ molecules with the ssDNA sequence 1 coated graphene.}
\label{fig.2}
\end{figure}

\begin{table*}[h!]
\caption{\label{tabone}Values for different model parameters used in the calculation of
the current with $\gamma_0$=3.3eV and $\varepsilon^\prime_f$=-0.7 eV for $\rm N_2$ with
DNA sequence 1.}
\label{log}
\begin{center}
\begin{tabular}{llllllll} \hline\hline \noalign{\smallskip}
$\bar{t}(s)$ & $\bar{t_1}$ & $\bar{t_2}$ & $\bar{t_3}$ & $\bar{t_4}$ & $\bar{t_5}$ & $\bar{t_6}$ & $\bar{t_7}$\\
\hline
$\Delta I^{(0)}(\bar{t})/{I_0}$ & $-2.5394$ & $-1.5962$ & $-1.2946$ & $-1.1855$ & $-0.9962$ & $-0.8542$ &$-0.5958$\\
$\varepsilon_f(\bar{t})(eV)$ ~& 1.77 ~& 1.71 ~& 1.6 ~& 1.59 ~& 1.55 ~& 1.52 ~& 1.50\\
${\rm \Delta a_{A_1B_1}}(\bar{t})$(\AA)&$0.02  $&$0.02  $&$0.02  $&$0.02  $&$0.02  $&$0.02  $&$0.02 $\\
${\rm \Delta a_{B_1A_2}}(\bar{t})$(\AA)&$-0.009$&$0.004 $&$0.004 $&$0.004 $&$0.004 $&$0.004 $&$0.004$ \\
${\rm \Delta a_{A_2B_2}}(\bar{t})$(\AA)&$-0.029$&$-0.018$&$0.022 $&$0.022 $&$0.022 $&$0.022 $&$0.022$\\
${\rm \Delta a_{B_2A_3}}(\bar{t})$(\AA)&$-0.03 $&$-0.03 $&$-0.01 $&$0.004 $&$0.004 $&$0.004 $&$0.004$\\
${\rm \Delta a_{A_3B_3}}(\bar{t})$(\AA)&$-0.019$&$-0.019$&$-0.019$&$-0.007$&$0.004 $&$0.004 $&$0.004$\\
${\rm \Delta a_{B_3A_4}}(\bar{t})$(\AA)&$-0.019$&$-0.019$&$-0.019$&$-0.019$&$-0.01 $&$0.004 $&$0.004$\\
${\rm \Delta a_{A_4B_4}}(\bar{t})$(\AA)&$-0.03 $&$-0.03 $&$-0.03 $&$-0.03 $&$-0.03 $&$-0.018$&$0.022$\\
${\rm \Delta a_{B_4A_5}}(\bar{t})$(\AA)&$-0.018$&$-0.018$&$-0.018$&$-0.018$&$-0.018$&$-0.018$&$0.002$\\
\noalign{\smallskip} \hline\hline
\noalign{\smallskip}
\end{tabular}
\end{center}
\end{table*}

\begin{figure}[h!]
\center
\includegraphics[width=10cm]{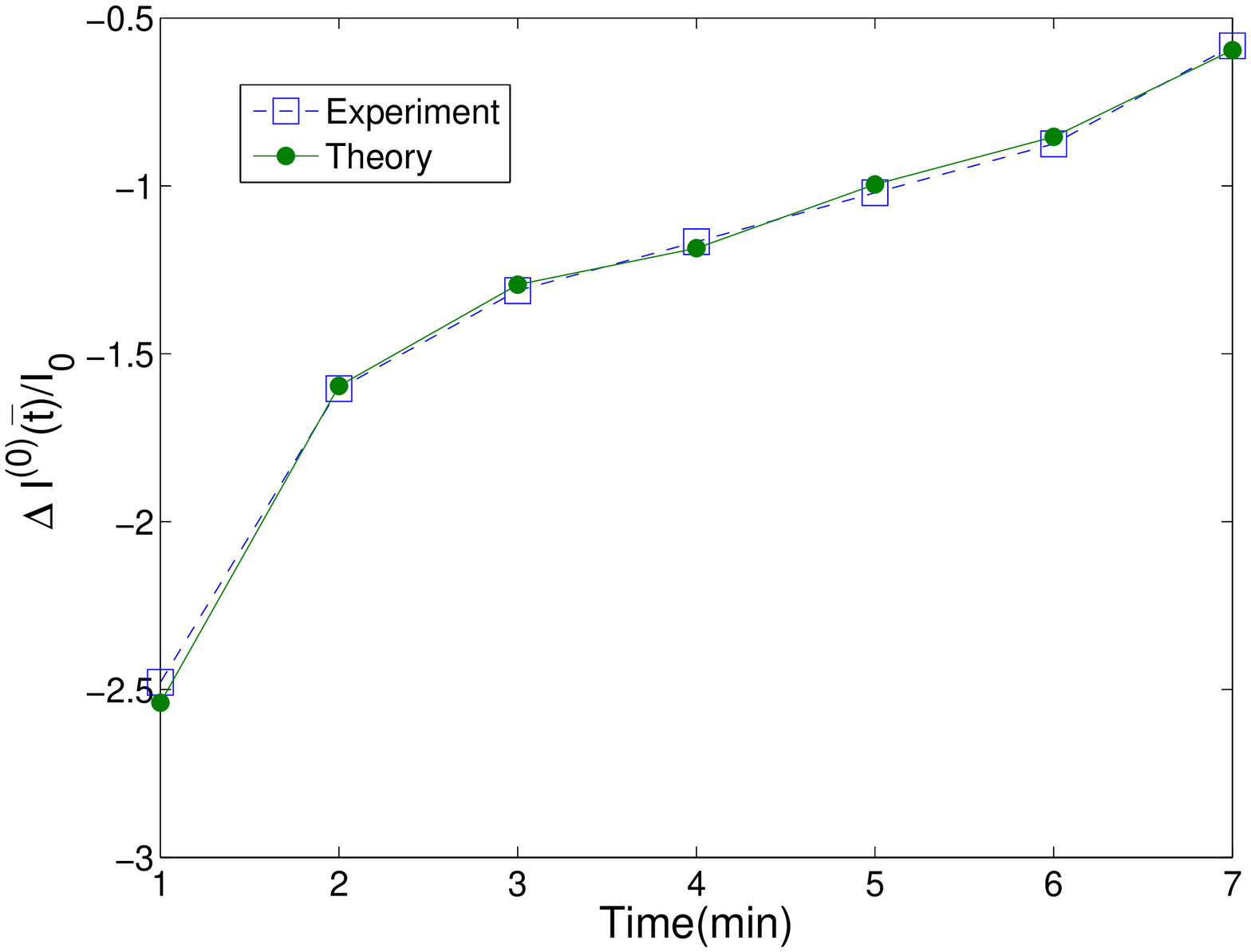}
\caption{The sensor response $\Delta I^{(0)}(\bar{t})/{I_0}$ versus exposure time $\bar{t}$(s)
for $\rm N_2$ with ssDNA sequence 1. Experimental data is from Ref. [37]. Theoretical and
experimental results are in agreement.}
\label{fig.2}
\end{figure}

\begin{table*}[h!]
\caption{\label{tabone}Values for different model parameters used in the calculation
of the current with $\gamma_0$=3.3eV and $\varepsilon^\prime_f$=-0.7eV for DMMP with
DNA sequence 2.}
\label{log}
\begin{center}
\begin{tabular}{llllllll} \hline\hline \noalign{\smallskip}
$\bar{t}(s)$ & $\bar{t_1}$ & $\bar{t_2}$ & $\bar{t_3}$ & $\bar{t_4}$ & $\bar{t_5}$ & $\bar{t_6}$ & $\bar{t_7}$\\
\hline
$\Delta I^{(0)}(\bar{t})/{I_0}$ & -2.2054 & -3.3501 & -3.8368 & -4.2537 & -4.5883 & -4.8608 & -4.9574\\
$\varepsilon_f(\bar{t})(eV)$  & 1.74 & 1.79 & 1.83 & 1.84 & 1.87 & 1.89 & 1.9\\
${\rm \Delta a_{A_1B_1}}(\bar{t})$(\AA)&-0.009&-0.009&-0.009&-0.009&-0.009&-0.009&-0.009\\
${\rm \Delta a_{B_1A_2}}(\bar{t})$(\AA)&-0.007&-0.019&-0.019&-0.019&-0.019&-0.019&-0.019\\
${\rm \Delta a_{A_2B_2}}(\bar{t})$(\AA)&---   &-0.01 &-0.022&-0.022&-0.022&-0.022&-0.022\\
${\rm \Delta a_{B_2A_3}}(\bar{t})$(\AA)&---   &---   &-0.01 &-0.019&-0.019&-0.019&-0.019\\
${\rm \Delta a_{A_3B_3}}(\bar{t})$(\AA)&---   &---   &---   &-0.007&-0.019&-0.019&-0.019\\
${\rm \Delta a_{B_3A_4}}(\bar{t})$(\AA)&---   &---   &---   &---   &-0.01 &-0.03 &-0.03\\
${\rm \Delta a_{A_4B_4}}(\bar{t})$(\AA)&---   &---   &---   &---   &---   &-0.018&-0.03\\
${\rm \Delta a_{B_4A_5}}(\bar{t})$(\AA)&---   &---   &---   &---   &---   &---   &-0.01\\
\noalign{\smallskip} \hline\hline
\noalign{\smallskip}
\end{tabular}
\end{center}
\end{table*}

\begin{figure}[h!]
\center
\includegraphics[width=10cm]{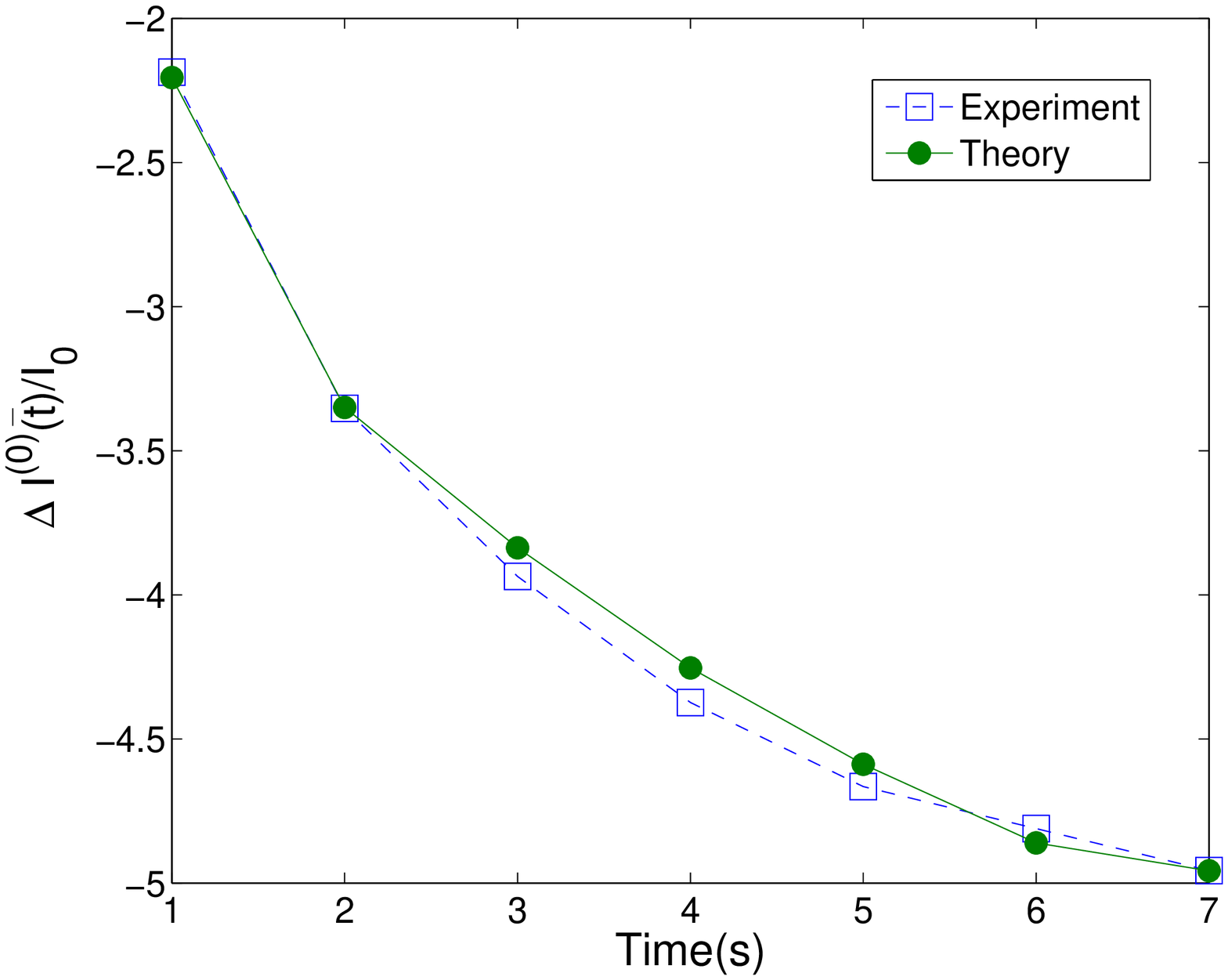}
\caption{The sensor response $\Delta I^{(0)}(\bar{t})/{I_0}$ versus exposure time
$\bar{t}$(s) for DMMP with ssDNA sequence 2. Theoretical and experimental results
are in agreement.}
\label{fig.2}
\end{figure}

\begin{table*}[h!]
\caption{\label{tabone}Values for different model parameters used in the calculation
of the current with $\gamma_0$=3.3eV and $\varepsilon^\prime_f$=-0.7eV for $\rm N_2$
with DNA sequence 2.}
\label{log}
\begin{center}
\begin{tabular}{llllllll} \hline\hline \noalign{\smallskip}
$\bar{t}(s)$&$\bar{t_1}$&$\bar{t_2}$&$\bar{t_3}$&$\bar{t_4}$&$\bar{t_5}$&$\bar{t_6}$&$\bar{t_7}$\\
\hline
$\Delta I^{(0)}(\bar{t})/{I_0}$ & -4.7486 & -4.4499 & -2.6233 & -2.3937 & -1.9362 & -1.5505 & -1.2979\\
$\varepsilon_f(\bar{t})(eV)$ & 1.88 & 1.87 & 1.78 & 1.77 & 1.72 & 1.64 & 1.61\\
${\rm \Delta a_{A_1B_1}}(\bar{t})$(\AA)&$0.002 $&$0.02  $&$0.02  $&$0.02 $&$0.02  $&$0.02 $&$0.02 $\\
${\rm \Delta a_{B_1A_2}}(\bar{t})$(\AA)&$-0.01 $&$0.004 $&$0.004 $&$0.004$&$0.004 $&$0.004$&$0.004$ \\
${\rm \Delta a_{A_2B_2}}(\bar{t})$(\AA)&$-0.022$&$-0.01 $&$0.004 $&$0.004$&$0.004 $&$0.004$&$0.004$\\
${\rm \Delta a_{B_2A_3}}(\bar{t})$(\AA)&$-0.019$&$-0.019$&$-0.007$&$0.004$&$0.004 $&$0.004$&$0.004$\\
${\rm \Delta a_{A_3B_3}}(\bar{t})$(\AA)&$-0.019$&$-0.019$&$-0.019$&$-0.01$&$0.004 $&$0.004$&$0.004$\\
${\rm \Delta a_{B_3A_4}}(\bar{t})$(\AA)&$-0.03 $&$-0.03 $&$-0.03 $&$-0.03$&$-0.018$&$0.022$&$0.022$\\
${\rm \Delta a_{A_4B_4}}(\bar{t})$(\AA)&$-0.03 $&$-0.03 $&$-0.03 $&$-0.03$&$-0.03 $&$-0.01$&$0.004$\\
${\rm \Delta a_{B_4A_5}}(\bar{t})$(\AA)&$-0.01 $&$-0.01 $&$-0.01 $&$-0.01$&$-0.01 $&$-0.01$&$0.002$\\
\noalign{\smallskip} \hline\hline
\noalign{\smallskip}
\end{tabular}
\end{center}
\end{table*}

\begin{figure}[h!]
\center
\includegraphics[width=10cm]{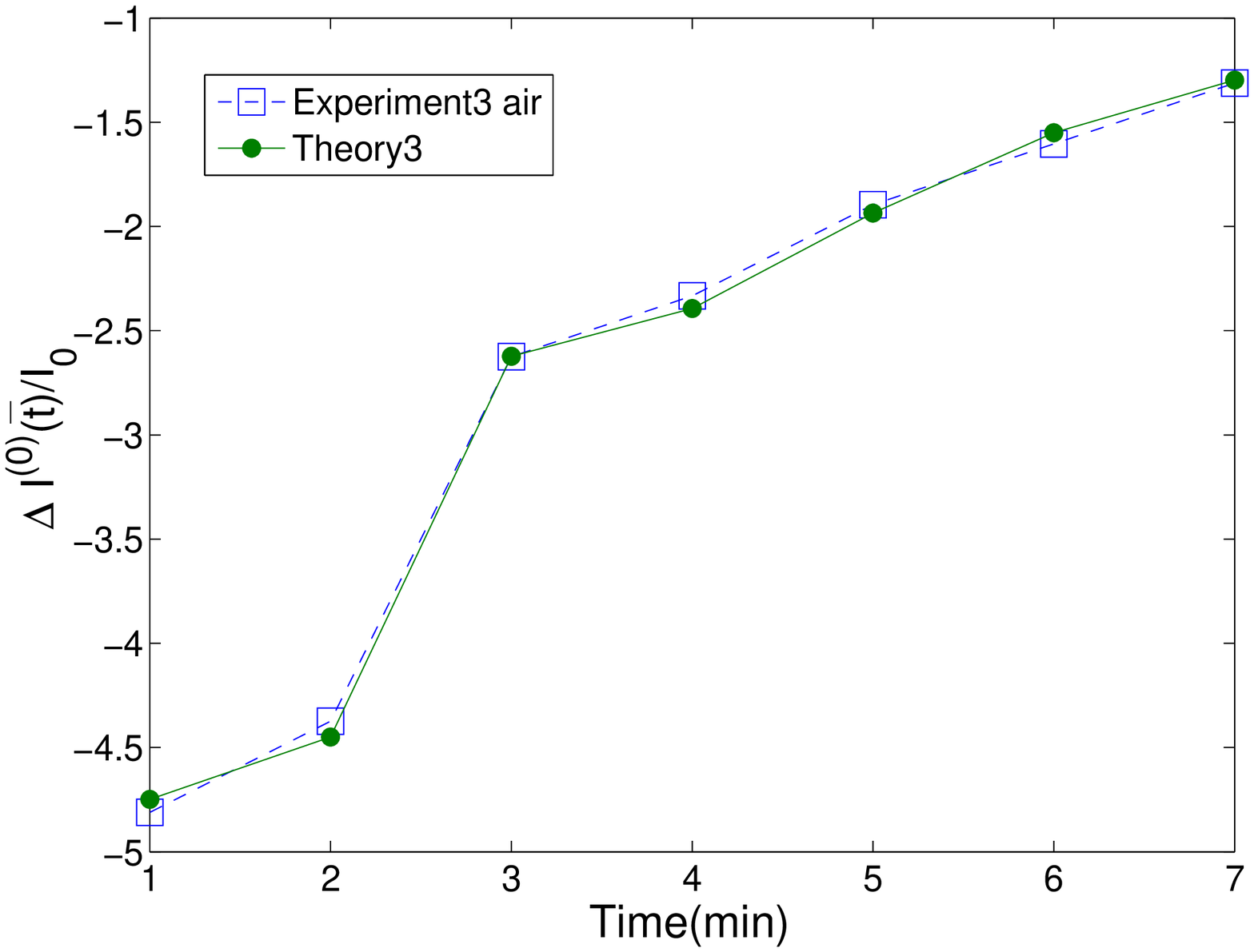}
\caption{The sensor response $\Delta I^{(0)}(\bar{t})/{I_0}$ versus exposure time
$\bar{t}$(s) for $\rm N_2$ with ssDNA sequence 2. Theoretical and experimental
results are in agreement.}
\label{fig.2}
\end{figure}

\section*{Acknowledgments}
This work has been supported by the Council of Scientific and Industrial Research
(CSIR), the University Grant Commission (UGC) and the University Faculty R $\&$ D
Research Programme.

%\newpage
\vskip 1in
\appendix
\section{Calculation of the sensor response}

\vskip 0.1in
The sensor response of graphene coated with ssDNA sequence 1 when exposed to DMMP
vapor is defined as
\begin{equation}
\Delta I^{(0)}(\bar t)/I_0=1-\frac{I^{(0)}(\bar t)({\rm vapor})}{I_0({\rm air})}=1-\frac{T(\varepsilon_f,\bar t)}{T_0
(\varepsilon_f^\prime,\bar t_{\rm max})},
\end{equation}

where $T(\varepsilon_f,\bar t)$ and $T_0 (\varepsilon_f^\prime,\bar t_{\rm max})$
are transmission functions of the DNA coated graphene when exposed to gas and
$\rm N_2$, respectively.

To calculate the sensor response we consider a simple $9\times 9$ matrix.

At time $\bar t_1$: the DMMP molecule interacts with the guanine base attached to the
carbon atom $\rm B_1$ causing charge transfer to graphene which leads to changes $\rm
\Delta a_{A_1B_1}(\bar t_1)$ and $\rm \Delta a_{B_1A_2}(\bar t_1)$ with the corresponding
hopping integrals ${\gamma_{11}(\bar{t_1})}$ and ${\gamma_{12}(\bar{t_1})}$ while the
other $\rm a_{cc}$ remain unaffected with hopping integrals $\gamma_0$. The transmission

\begin{equation}
\frac{T(\varepsilon_f,\bar t_1)}{T_0(\varepsilon_f^\prime, \bar t_{\rm max})}=
\frac{\frac{4~{\rm Im}{\Sigma^r_{\rm S}}~{\rm Im}{\Sigma^r_{\rm D}}~{\gamma^{2}_{11}(\bar{t_1})}
{\gamma^{2}_{12}(\bar{t_1})}{\gamma^{2}_{0}}{\gamma^{2}_{0}}{\gamma^{2}_{0}}
{\gamma^{2}_{0}}{\gamma^{2}_{0}}{\gamma^{2}_{{0}}}}{|{G^{(0)r}_{\rm grp}}
(\varepsilon_f,\bar{t_1})|^2_{9\times9}}}{\frac{4~{\rm Im}{\Sigma^r_{\rm S}}~{\rm Im}
{\Sigma^r_{\rm D}}~{\gamma^{2}_{0}}{\gamma^{2}_{0}}{\gamma^{2}_{0}}{\gamma^{2}_{0}}{\gamma^{2}_{0}}
{\gamma^{2}_{0}}{\gamma^{2}_{0}}{\gamma^{2}_{0}}}{|{G^{(0)r}_{\rm grp}}(\varepsilon_f^\prime, \bar
t_{\rm max})|^2_{9 \times 9}}},
\end{equation}

where

\begin{eqnarray}
|G_{\rm grp}^{(0)r}(\varepsilon_f,\bar{t_1})|&=&\varepsilon_f^9(\bar{t_1})-\varepsilon_f^7(\bar{t_1})\gamma^2_{11}(\bar{t_1})-\varepsilon_f^7(\bar{t_1})
\gamma^2_{12}(\bar{t_1})-\varepsilon_f^7(\bar{t_1})\gamma^2_{0}+\varepsilon_f^5(\bar{t_1})\gamma^2_{11}(\bar{t_1})\gamma^2_{0}
\nonumber\\
&-&\varepsilon_f^7(\bar{t_1})\gamma^2_{0}+\varepsilon_f^5(\bar{t_1})\gamma^2_{11}(\bar{t_1})\gamma^2_{0}+\varepsilon_f^5(\bar{t_1})
\gamma^2_{12}(\bar{t_1})\gamma^2_{0}-\varepsilon_f^7(\bar{t_1})\gamma^2_{0}\nonumber\\
&+&\varepsilon_f^5(\bar{t_1})\gamma^2_{11}(\bar{t_1})\gamma^2_{0}+\varepsilon_f^5(\bar{t_1})\gamma^2_{12}(\bar{t_1})\gamma^2_{0}
+\varepsilon_f^5(\bar{t_1})\gamma^2_{0}\gamma^2_{0}\nonumber\\
&-&\varepsilon_f^3(\bar{t_1})\gamma^2_{11}(\bar{t_1})\gamma^2_{0}\gamma^2_{0}-\varepsilon_f^7(\bar{t_1})\gamma^2_{0}
+\varepsilon_f^5(\bar{t_1})\gamma^2_{11}(\bar{t_1})\gamma^2_{0}\nonumber\\
&+&\varepsilon_f^5(\bar{t_1})\gamma^2_{12}(\bar{t_1})\gamma^2_{0}+\varepsilon_f^5(\bar{t_1})\gamma^2_{0}\gamma^2_{0}
-\varepsilon_f^3(\bar{t_1})\gamma^2_{11}(\bar{t_1})\gamma^2_{0}\gamma^2_{0}\nonumber\\
&+&\varepsilon_f^5(\bar{t_1})\gamma^2_{0}\gamma^2_{0}-\varepsilon_f^3(\bar{t_1})\gamma^2_{11}(\bar{t_1})\gamma^2_{0}
\gamma^2_{0}-\varepsilon_f^3(\bar{t_1})\gamma^2_{12}(\bar{t_1})\gamma^2_{0}\gamma^2_{0}\nonumber\\
&-&\varepsilon_f^7(\bar{t_1})\gamma^2_{0}+\varepsilon_f^5(\bar{t_1})\gamma^2_{11}(\bar{t_1})\gamma^2_{0}+\varepsilon_f^5(\bar{t_1})
\gamma^2_{12}(\bar{t_1})\gamma^2_{0}+\varepsilon_f^5(\bar{t_1})\gamma^2_{0}\gamma^2_{0}\nonumber\\
&-&\varepsilon_f^3(\bar{t_1})\gamma^2_{11}(\bar{t_1})\gamma^2_{0}\gamma^2_{0}
+\varepsilon_f^5(\bar{t_1})\gamma^2_{0}\gamma^2_{0}-\varepsilon_f^3(\bar{t_1})\gamma^2_{11}(\bar{t_1})\gamma^2_{0}
\gamma^2_{0}\nonumber\\
&-&\varepsilon_f^3(\bar{t_1})\gamma^2_{12}(\bar{t_1})\gamma^2_{0}\gamma^2_{0}+\varepsilon_f^5(\bar{t_1})\gamma^2_0
\gamma^2_{0}-\varepsilon_f^3(\bar{t_1})\gamma^2_{11}(\bar{t_1})\gamma^2_{0}\gamma^2_{0}\nonumber\\
&-&\varepsilon_f^3(\bar{t_1})\gamma^2_{12}(\bar{t_1})\gamma^2_{0}\gamma^2_{0}-\varepsilon_f^3(\bar{t_1})\gamma^2_{0}
\gamma^2_{0}\gamma^2_{0}\nonumber\\
&+&\varepsilon_f(\bar{t_1})\gamma^2_{11}(\bar{t_1})\gamma^2_{0}\gamma^2_{0}\gamma^2_{0}-\varepsilon_f^7(\bar{t_1})
\gamma^2_{0}+\varepsilon_f^5(\bar{t_1})\gamma^2_{11}(\bar{t_1})\gamma^2_{0}\nonumber\\
&+&\varepsilon_f^5(\bar{t_1})\gamma^2_{12}(\bar{t_1})\gamma^2_{0}+\varepsilon_f^5(\bar{t_1})\gamma^2_{0}(\bar{t_1})
\gamma^2_{0}-\varepsilon_f^3(\bar{t_1})\gamma^2_{11}(\bar{t_1})\gamma^2_{0}\gamma^2_{0}\nonumber\\
&+&\varepsilon_f^5(\bar{t_1})\gamma^2_{0}\gamma^2_{0}-\varepsilon_f^3(\bar{t_1})\gamma^2_{11}(\bar{t_1})\gamma^2_{0}
\gamma^2_{0}-\varepsilon_f^3(\bar{t_1})\gamma^2_{12}(\bar{t_1})\gamma^2_{0}\gamma^2_{0}\nonumber\\
&+&\varepsilon_f^5(\bar{t_1})\gamma^2_{0}\gamma^2_{0}-\varepsilon_f^3(\bar{t_1})\gamma^2_{11}(\bar{t_1})\gamma^2_{0}
\gamma^2_{0}-\varepsilon_f^3(\bar{t_1})\gamma^2_{12}(\bar{t_1})\gamma^2_{0}\gamma^2_{0}\nonumber\\
&-&\varepsilon_f^3(\bar{t_1})\gamma^2_{0}\gamma^2_{0}\gamma^2_{0}+\varepsilon_f(\bar{t_1})\gamma^2_{11}(\bar{t_1})
\gamma^2_{0}\gamma^2_{0}\gamma^2_{0}+\varepsilon_f^5(\bar{t_1})\gamma^2_{0}\gamma^2_{0}\nonumber\\
&-&\varepsilon_f^3(\bar{t_1})\gamma^2_{11}(\bar{t_1})\gamma^2_{0}\gamma^2_{0}-\varepsilon_f^3(\bar{t_1})\gamma^2_{12}(\bar{t_1})
\gamma^2_{0}\gamma^2_{0}-\varepsilon_f^3(\bar{t_1})\gamma^2_{0}\gamma^2_{0}\gamma^2_{0}\nonumber\\
&+&\varepsilon_f(\bar{t_1})\gamma^2_{11}(\bar{t_1})\gamma^2_{0}\gamma^2_{0}\gamma^2_{0}-\varepsilon_f^3(\bar{t_1})
\gamma^2_{0}\gamma^2_{0}\gamma^2_{0}\nonumber\\
&+&\varepsilon_f(\bar{t_1})\gamma^2_{11}(\bar{t_1})\gamma^2_{0}\gamma^2_{0}\gamma^2_{0}+\varepsilon_f(\bar{t_1})
\gamma^2_{12}(\bar{t_1})\gamma^2_{0}\gamma^2_{0}\gamma^2_{0},
\end{eqnarray}

and

\begin{equation}
|{G^{(0)r}_{\rm grp}}(\varepsilon_f^\prime,\bar t_{\rm max})|=\varepsilon_f^{\prime^9}-8\varepsilon_f^{\prime^7}\gamma_0^2 +21\varepsilon_f^{\prime^5}\gamma_0^4-20\varepsilon_f^{\prime^3}\gamma_0^6+5\varepsilon_f^\prime\gamma_0^8.
\end{equation}

Here we have neglected the contribution from the on-site energy as we are interested in
the transport properties (hopping), and the energy independent self-energy terms are
also ignored for simplification of the calculation.

Using the form for the hopping integral $\gamma_{\rm ij}(\Delta {\rm a_{cc}}(\bar{t}))= \gamma_0 {\rm exp(-\Delta a_{cc}}(\bar{t})/{\rm a_0})$
the transmission function ratio becomes

\begin{equation}
\frac{T(\varepsilon_f,\bar t_1)}{T_0(\varepsilon_f^\prime, \bar t_{\rm max})}=
\frac{\frac{4~{\rm Im}{\Sigma^r_{\rm S}}~{\rm Im}{\Sigma^r_{\rm D}}~{\gamma^{2}_{0}[{\rm exp}(-{\rm \Delta a_{A1B1}}(\bar{t_1})/{\rm a_0})]^2}
{\gamma^{2}_{0}[{\rm exp}( -{\rm\Delta a_{B1A2}}(\bar{t_1})/{\rm a_0})]^2}{\gamma^{2}_{0}}{\gamma^{2}_{0}}{\gamma^{2}_{0}}
{\gamma^{2}_{0}}{\gamma^{2}_{0}}{\gamma^{2}_{{0}}}}{|{G^{(0)r}_{\rm grp}}
(\varepsilon_f,\bar{t_1})|^2_{9\times9}}}{\frac{4~{\rm Im}{\Sigma^r_{\rm S}}~{\rm Im}
{\Sigma^r_{\rm D}}~{\gamma^{2}_{0}}{\gamma^{2}_{0}}{\gamma^{2}_{0}}{\gamma^{2}_{0}}{\gamma^{2}_{0}}
{\gamma^{2}_{0}}{\gamma^{2}_{0}}{\gamma^{2}_{0}}}{|{G^{(0)r}_{\rm grp}}(\varepsilon_f^\prime, \bar
t_{\rm max})|^2_{9 \times 9}}}.
\end{equation}

Equations (A3) and (A5) give two model parameters ${\rm \Delta a_{cc}}(\bar t)$ and $\varepsilon_f(\bar t)$.
To calculate the current we find the values of these two parameters.

The following steps are used to calculate the sensor response:

\begin{itemize}
\item
The constants $\gamma_0$, $\varepsilon_f^\prime(\bar t_{\rm max})$ and $\rm a_0$ are set as 3.3 eV, -0.7eV,
and 0.33 \AA, respectively, which give the best matching between the theoretical and experimental results.

\item A range of the values of the Fermi level $\varepsilon_f(\bar t)$ is set from 0.01 to 4 with an interval of 0.01.

\item \textbf{At time $\bar t_1$:} the gas molecule interacts with the guanine base causing changes ${\rm \Delta a_{A_1B_1}}(\bar t_1)$
and ${\rm \Delta a_{B_1A_2}}(\bar t_1)$.
%with corresponding hopping integrals $\gamma_{11}(\bar t_1)$ and $\gamma_{12}(\bar t_1)$.
The best fitted values of ${\rm \Delta a_{A_1B_1}}(\bar t_1)$ and ${\rm \Delta a_{B_1A_2}}(\bar t_1)$ due to guanine base are

\vskip 0.1in
\noindent
${\rm \Delta a_{A_1B_1}}(\bar t_1)=-0.02$
\vskip 0.1in
\noindent
${\rm \Delta a_{B_1A_2}}(\bar t_1)=-0.018$

\item
Substituting these values in the transmission function ratio, Eq. (A5), and using
Eq. (A1)
%$\Delta I^{(0)}(\bar t)/I_0=1-\frac{T(\varepsilon_f,\bar t)}{T_0(\varepsilon_f^\prime,\bar t_{\rm max})}$
the sensor response is calculated.

\item
To read the values of the current with corresponding values of $\varepsilon_f (\bar t_1)$ explicitly, a Matrix is defined as

$A=\{\varepsilon_f (\bar t_1),\Delta I^{(0)}(\bar t_1)/I_0\}$

\newpage
\item
To read the elements of the matrix A the following code is typed

%\vskip 0.1in
%\noindent
%A$[All,\{1,2,3,4,5,6,7,8,9,10}]//$MatrixForm
\begin{figure}[h!]
%\vspace {-1cm}
%\vspace {-0.8cm}
\begin{center}
\includegraphics[width=8cm]{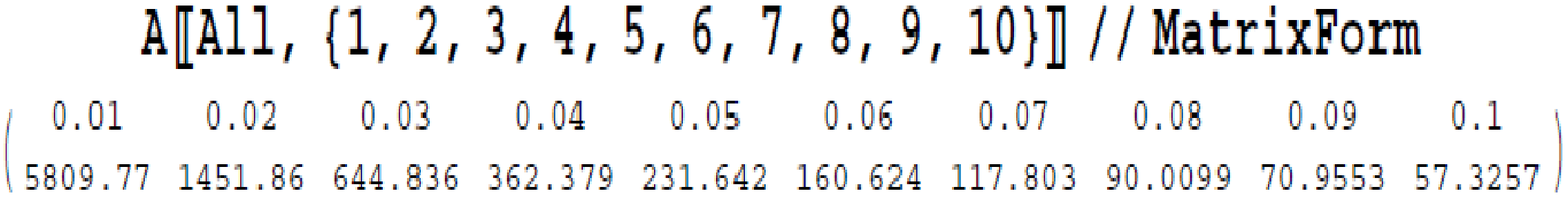}
\end{center}
\end{figure}
%\vspace {-1cm}
where the upper row shows the values of $\varepsilon_f (\bar t_1)$ and the lower row shows the
corresponding values of the sensor response.

\item
Similarly other matrix elements are read as
\begin{figure}[h!]
%\vspace {-1.2cm}
\begin{center}
\includegraphics[width=8cm]{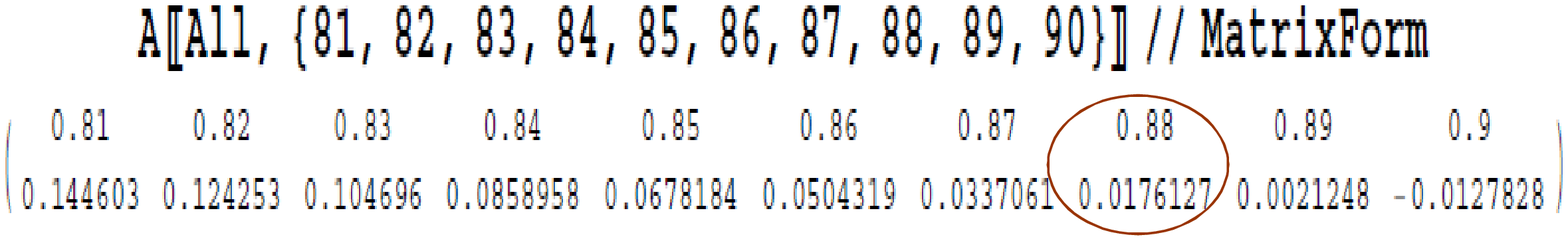}
\end{center}
\end{figure}

where the sensor response $\Delta I^{(0)}(\bar t_1)/I_0$=0.0176127 at $\varepsilon_f (\bar t_1)$=0.88
highlighted with a circle is the theoretical sensor response which matches with the experimental
result very well.

\item
\textbf{At time $\bar t_2$:} the DMMP molecule interacts with the adenine base attached to the carbon atom $\rm A_2$
leading to changes ${\rm \Delta a_{B_1A_2}}(\bar t_2)$ and ${\rm \Delta a_{A_2B_2}}(\bar t_2)$ with
corresponding hopping integrals $\gamma_{12}(\bar t_2)$ and $\gamma_{22}(\bar t_2)$, whereas ${\rm \Delta a_{A_1B_1}}(\bar t_2)$
remains unaffected. The best fitted values of ${\rm \Delta a_{B_1A_2}}(\bar t_2)$ and ${\rm \Delta a_{A_2B_2}}(\bar t_2)$
due to adenine base are

\vskip 0.1in
\noindent
${\rm \Delta a_{A_1B_1}}(\bar t_2)=-0.02$
\vskip 0.1in
\noindent
${\rm \Delta a_{B_1A_2}}(\bar t_2)=-0.018-0.011=-0.029$
\vskip 0.1in
\noindent
${\rm \Delta a_{B_1A_2}}(\bar t_2)=-0.009$
%\end{itemize}

Similarly, the sensor response is calculated at next instants of time.

\item
\textbf{At $\bar t_3$:} the changes in $\rm \Delta a_{cc}$ due to guanine base are

${\rm \Delta a_{A1B1}}(\bar{t_3})=-0.02$

${\rm \Delta a_{B1A2}}(\bar{t_3})=-0.029$

${\rm \Delta a_{A2B2}}(\bar{t_3})=-0.009-0.02=-0.029$

${\rm \Delta a_{B2A3}}(\bar{t_3})=-0.018$

\item
\textbf{At $\bar t_4$:} the changes in $\rm \Delta a_{cc}$ due to thymine base are

${\rm \Delta a_{A1B1}}(\bar{t_4})=-0.02$

${\rm \Delta a_{B1A2}}(\bar{t_4})=-0.029$

${\rm \Delta a_{A2B2}}(\bar{t_4})=-0.029$

${\rm \Delta a_{B2A3}}(\bar{t_4})=-0.018-0.012=-0.03$

${\rm \Delta a_{A3B3}}(\bar{t_4})=-0.01$

\item
\textbf{At $\bar t_5$:} the changes in $\rm \Delta a_{cc}$ due to cytosine base are

${\rm \Delta a_{A1B1}}(\bar{t_5})=-0.02$

${\rm \Delta a_{B1A2}}(\bar{t_5})=-0.029$

${\rm \Delta a_{A2B2}}(\bar{t_5})=-0.029$

${\rm \Delta a_{B2A3}}(\bar{t_5})=-0.03$

${\rm \Delta a_{A3B3}}(\bar{t_5})=-0.01-0.009=-0.019$

${\rm \Delta a_{B3A4}}(\bar{t_5})=-0.007$

\item
\textbf{At $\bar t_6$:} the changes in $\rm \Delta a_{cc}$ due to thymine base are

${\rm \Delta a_{A1B1}}(\bar{t_6})=-0.02$

${\rm \Delta a_{B1A2}}(\bar{t_6})=-0.029$

${\rm \Delta a_{A2B2}}(\bar{t_6})=-0.029$

${\rm \Delta a_{B2A3}}(\bar{t_6})=-0.03$

${\rm \Delta a_{A3B3}}(\bar{t_6})=-0.019$

${\rm \Delta a_{B3A4}}(\bar{t_6})=-0.007-0.012=-0.019$

${\rm \Delta a_{A4B4}}(\bar{t_6})=-0.01$

\item
\textbf{At $\bar t_7$:} the changes in $\rm \Delta a_{cc}$ due to guanine base are

${\rm \Delta a_{A1B1}}(\bar{t_7})=-0.02$

${\rm \Delta a_{B1A2}}(\bar{t_7})=-0.029$

${\rm \Delta a_{A2B2}}(\bar{t_7})=-0.029$

${\rm \Delta a_{B2A3}}(\bar{t_7})=-0.03$

${\rm \Delta a_{A3B3}}(\bar{t_7})=-0.019$

${\rm \Delta a_{B3A4}}(\bar{t_7})=-0.019$

${\rm \Delta a_{A4B4}}(\bar{t_7})=-0.01-0.02=-0.03$

${\rm \Delta a_{B4A5}}(\bar{t_7})=-0.018$
\end{itemize}

Similar steps are followed for the calculation of the sensor response
when graphene coated with ssDNA sequence 2 is exposed to DMMP and $\rm N_2$.

\bibliographystyle{model1a-num-names}
\bibliography{<your-bib-database>}

\end{document}